\documentclass[twocolumn]{aastex631}

\usepackage{longtable,morefloats}
\usepackage{apjfonts}
\usepackage{mathrsfs}
\usepackage{booktabs}
\usepackage{hyperref}
\usepackage{amsmath}
\usepackage{natbib}
\usepackage{bm}
\usepackage{subfigure}
\usepackage{multirow}
\usepackage{gensymb}

\newcommand{\xmm}{\hbox{XMM-Newton\/}}
\newcommand{\chandra}{{Chandra\/}}
\newcommand{\xray}{{\hbox{X-ray}}}

\newcommand\iona[2]{#1$\;${\scshape{#2}}}

\begin{document}

\title{Rest-frame Optical Spectroscopy of $z \sim 2$ Quasars with Steep Hard X-ray Spectral Shapes: X-ray Selection of Super-Eddington Accretion and Verification}
\author{Y.~Chen}
\affiliation{School of Astronomy and Space Science, Nanjing University, Nanjing, Jiangsu 210093, China; bluo@nju.edu.cn}
\affiliation{Key Laboratory of Modern Astronomy and Astrophysics (Nanjing University), Ministry of Education, Nanjing 210093, China}

\author{B.~Luo}
\affiliation{School of Astronomy and Space Science, Nanjing University, Nanjing, Jiangsu 210093, China; bluo@nju.edu.cn}
\affiliation{Key Laboratory of Modern Astronomy and Astrophysics (Nanjing University), Ministry of Education, Nanjing 210093, China}

\author{J.~Huang}
\affiliation{School of Astronomy and Space Science, Nanjing University, Nanjing, Jiangsu 210093, China; bluo@nju.edu.cn}
\affiliation{Key Laboratory of Modern Astronomy and Astrophysics (Nanjing University), Ministry of Education, Nanjing 210093, China}


%%%%%%%%%%%%%%%%%%%%%%%%%%%%
\begin{abstract}

Super-Eddington accretion is a crucial phase in the growth of supermassive black holes.
However, identifying super-Eddington accreting quasars observationally is challenging due to uncertain black-hole mass estimates and other complications.
The Eddington ratio parameter does not represent accurately the accretion rate in the super-Eddington regime.
On the other hand, super-Eddington accreting quasars appear to show large hard X-ray (rest-frame $>$ 2 keV) power-law photon indices,
and they also exhibit distinct optical spectral features including weak [\iona{O}{iii}]~$\lambda 5007$ emission and strong \iona{Fe}{ii} emission.
We test steep X-ray photon-index selection of super-Eddington accreting quasars by obtaining Palomar 200-inch Hale Telescope near-infrared spectra for a pilot sample of nine $\Gamma=2.0$--2.6 quasars at $z\approx1.4$--2.5.
We derive H$\beta$-based single-epoch virial black-hole masses (median value $\rm 4.3 \times 10^{8}~M_{\odot}$) and Eddington ratios (median value 0.6).
The Eddington ratio distribution is consistent with that of the comparison sample, which is a flux-limited sample of quasars at $z\approx1.5$--3.5 with near-infrared spectroscopy.
But our super-Eddington candidates do show statistically weaker [\iona{O}{iii}] emission ($P_{\rm null}=0.0075$) and marginally stronger \iona{Fe}{ii} emission ($P_{\rm null}=0.06$).
We also find one candidate with broad (width of 1960 km/s) and blueshifted (690 km/s) [\iona{O}{iii}]~$\lambda 4959$ and [\iona{O}{iii}]~$\lambda 5007$ lines, which probably originate from a strong [\iona{O}{iii}] outflow driven by super-Eddington accretion.
Overall, the steep X-ray photon-index selection of super-Eddington accreting quasars appears promising.
But a larger sample is needed to assess further the reliability of the selection.
\end{abstract}

%%%%%%%%%%%%%%%%%%%%%%%%%%%%
\section{Introduction}\label{sec:intro}

Quasars are powered by accretion of fueling gas onto supermassive black holes (SMBHs).
A large fraction of the quasar power is released in the optical-to-X-ray energies.
The accretion disk generates thermal radiation in the optical and ultraviolet (UV; e.g., \citealt{Shields1978}), while hot elections in the accretion disk corona are able to inverse Compton scatter the optical/UV photons into the X-ray energies \citep[e.g.,][]{Haardt1991}.
Besides powering the quasar multiwavelength radiation, accretion is likely the main growth channel of SMBHs \citep[e.g.,][]{Soltan1982,Marconi2004,Zou2024}.
Discoveries of luminous quasars hosting massive SMBHs in the early universe \citep[e.g.,][]{Wu2015} pose challenges to our understanding of SMBH assembly, and super-Eddington accretion\footnote{The accretion rate exceeds the Eddington limit.
It is usually quantified by the Eddington ratio parameter, $\lambda_{\rm Edd}=L_{\rm Bol}/L_{\rm Edd}$, where $L_{\rm Bol}$ is the bolometric luminosity and $L_{\rm Edd}$ is the Eddington luminosity; $L_{\rm Edd}=1.5\times 10^{38}~M_{\rm BH}/{\rm M_{\odot}}$, where $M_{\rm BH}$ is the SMBH mass.} is generally invoked to explain rapid SMBH growth in the high-redshift universe (e.g., \citealt{Inayoshi2020}, and references therein).
Recent James Webb Space Telescope (JWST) discoveries of unexpected large population of high-redshift ($z>4$) active galactic nuclei (AGNs) likely also point to super-Eddington accretion for growing these massive black holes \citep[e.g.,][]{Akins2024,Greene2024,Kocevski2024,Maiolino2024,Madau2025}.
In addition, super-Eddington accretion in quasars may drive powerful outflows that could impact star-forming processes in their host galaxies \citep[e.g.,][]{Giustini2019,Jiang2019}, contributing to the SMBH-galaxy coevolution scenario.

Despite the importance of super-Eddington accretion, it is very challenging to identify super-Eddington accreting quasars via the Eddington ratio parameter, as the SMBH mass and bolometric luminosity measurements have substantial uncertainties. 
The single-epoch virial SMBH masses typically have $\approx0.4$--0.5 dex systematic uncertainties \citep[e.g.,][]{Shen2013} and $\approx0.15$ dex measurement uncertainties \citep[e.g.,][]{Shen2011}.
The empirical broad line region (BLR) radius--luminosity ($R$--$L$) relation appears to break down in the super-Eddington regime \citep[e.g.,][]{Hu2008,Wang20142,Du2015,Du2016,gravity2024},
and thus single-epoch virial SMBH masses would be biased high to levels that depend on the accretion rate.
Moreover, the virial assumption might even not be valid for super-Eddington accretion due to the impact of large radiation pressure and anisotropy of the ionizing radiation \citep[e.g.,][]{Marconi2008,Marconi2009,Netzer2010,Krause2011,Pancoast2014,Li2018}.
There are also significant systematic  uncertainties and/or biases (biased low) associated with $L_{\rm Bol}$ when a lot of the radiation is expected to be released in the poorly probed extreme UV (EUV; $\sim100$--$1200~\textup{\AA}$) for super-Eddington accretion \citep[e.g.,][]{Wang2014,Castell2016,Kubota2018}.
In addition, the $L_{\rm Bol}$ parameter (and thus $\lambda_{\rm Edd}$) is probably not a good representative of the accretion power in the super-Eddington regime, as a large fraction of the energy may be advected into the SMBH or be converted into the mechanical power of wind \citep[e.g.,][]{Wang2014,Jiang2019}.
The above complications significantly limit observational selection and characterization of super-Eddington accreting AGNs.
And because of these, when selecting super-Eddington accreting AGNs, the $\lambda_{\rm Edd}$ threshold is often not set at unity, but at lower values (e.g., $\lambda_{\rm Edd}\gtrsim0.3$; \citealt{Wang20142,Du2016}).

Previous studies have found a significant positive correlation between the hard X-ray ($>$ 2 keV) power-law photon index ($\Gamma$) and the Eddington ratio for typical AGNs \citep[e.g.,][]{Shemmer2008,Brightman2013}, 
which appears to extend to the super-Eddington regime \citep[e.g.,][]{Huang2020,Liu2021}.
Specifically, \cite{Liu2021} used a sample of 47 local AGNs with reverberation mapping SMBH mass measurements, including 21 super-Eddington accreting AGNs and 26 sub-Eddington accreting AGNs separated at $\lambda_{\rm Edd}\approx0.3$.
Statistically significant linear relations between $\Gamma$ and ${\rm log}(\lambda_{\rm Edd})$ were found for both the full sample and the super-Eddington sample, with intercepts of $\approx2.15$ (i.e., $\Gamma\approx2.15$ for $\lambda_{\rm Edd}=1$).
For $\lambda_{\rm Edd}=0.3$, the relations predict $\Gamma\approx2.0$.
The physics behind this empirical correlation is not clear.
One typically adopted explanation is that the X-ray corona cools more efficiently when the accretion rate is higher, which leads to a lower temperature and/or a smaller optical depth of the corona, yielding fewer hard X-ray photons \citep[e.g.,][]{Fabian2015,Yang2015,Kara2017,Ricci2018,Barua2020}.
We note that the $\Gamma$--$\lambda_{\rm Edd}$ relations generally exhibit large scatter, and there are also several studies claiming weak $\Gamma$--$\lambda_{\rm Edd}$ correlations \citep[e.g.,][]{Trakhtenbrot2017,Zhu2021,Kamraj2022,Laurenti2022,Laurenti2024,Trefoloni2023}. 
The complications may arise from multiple factors, including the accuracy of $\lambda_{\rm Edd}$ (both $M_{\rm BH}$ and $L_{\rm Bol}$), the dynamical range of $\lambda_{\rm Edd}$, X-ray data quality, details of X-ray spectral modeling, contamination from X-ray absorbed quasars, and X-ray differences between radio-loud and radio-quiet AGNs.

Nevertheless, steep X-ray spectral shapes ($\Gamma>2.0$) appear to be an alternative criterion for selecting super-Eddington accreting AGNs.
Currently, the public Chandra and XMM-Newton source catalogs \citep[e.g.,][]{Evans2010,Webb2020} cover $\approx5\%$ of the entire sky and provide X-ray measurements for large samples of AGNs, allowing in principle quick selection of super-Eddington accreting AGN candidates.
One complication in the X-ray selection is that the lower energy bounds for Chandra and XMM-Newton are around 0.3--0.5~keV, and for low-redshift AGNs, the spectral shape might be significantly contaminated by the frequent soft excess component that dominates the $\lesssim1$~keV emission, yielding many spurious super-Eddington candidates.
Therefore, such a catalog selection technique is best applicable to $z\gtrsim2$ quasars where the cataloged $\Gamma$ measurements are for the rest-frame $\gtrsim1$~keV energies and are more representative of the true coronal spectral shapes, significantly reducing the workload in the following spectral analysis that is needed to verify the large photon indices.\footnote{The Chandra and XMM-Newton source catalogs were generated using primarily photometric approaches in the observed frame with additional assumptions (e.g., $\Gamma = 1.7$ was assumed for converting XMM-Newton counts to fluxes; \citealt{Rosen2016}). 
Therefore, follow-up spectral fitting is always required to verify the rest-frame $\rm >2~keV$ $\Gamma$ value.}

Super-Eddington accreting quasars also display remarkable X-ray emission strengths, showing much higher occurrence rates of extreme X-ray weakness and X-ray variability \citep[e.g.,][]{Liu2019,Liu2021,Laurenti2022,Ni2022} compared to typical quasars \citep[e.g.,][]{Pu2020,Timlin2020}.
They may vary between X-ray nominal-strength states and X-ray weak states without contemporaneous optical/infrared variability, which hints that the intrinsic coronal emission is stable while the observed X-rays are modified by variable absorption.
The timescales for such extreme X-ray variability range from years down to days \citep[e.g.,][]{Liu2022}, pointing to a small-scale X-ray absorber.
Another unusual property is that some of these quasars show no absorption signatures in their X-ray weak states (i.e., displaying a steep power-law spectral shape), 
leading to the proposal that they are intrinsically X-ray weak rather than being affected by absorption \citep[e.g.,][]{Leighly2007,Laurenti2022,Trefoloni2023}.
However, Compton-thick absorption combined with a small leaked/scattered fraction of the intrinsic power-law continuum is also a viable explanation in some cases, which requires high-quality hard X-ray data to discern \citep{Wang2022}.
It is also worth noting that the prototypical intrinsically X-ray weak quasar, PHL 1811, was observed to briefly recover to the X-ray normal state \citep{Li2024}, which is consistent with the obscuration scenario.
Although selection of super-Eddington accretion via extreme X-ray weakness or X-ray variability properties is not straightforward, available multi-epoch flux measurements from the Chandra and XMM-Newton source catalogs allow quick assessment of these features for any selected candidates.

Besides the above X-ray characteristics, a few optical spectral features have been proposed to be associated with super-Eddington accretion, which originate from the ``Eigenvector 1’’ (EV1) correlations \citep[e.g.,][]{Boroson1992, Sulentic2000, Shen2014}.
Super-Eddington accreting AGNs tend to exhibit weak [\iona{O}{iii}]~$\lambda5007$ emission and strong optical \iona{Fe}{ii} emission, quantified with small [\iona{O}{iii}] rest-frame equivalent widths (REWs) and large $R_{\rm Fe\ II}$\footnote{The relative strength of the optical \iona{Fe}{ii} emission in the rest-frame 4434--4684~\AA\ band to the broad H$\beta$ emission, ${\rm EW_{\rm Fe\ II}}/{\rm EW_{\rm H\beta}}$, where ${\rm EW_{\rm Fe\ II}}$ and ${\rm EW_{\rm H\beta}}$ are the REWs of the \iona{Fe}{ii} and broad H$\beta$ lines, respectively.} parameters.
The physical nature for these features remains unclear (see discussion in Section~4.1 of \citealt{Chen2024}), but empirically they do offer an alternative approach to select or cross-verify super-Eddington accreting AGNs.
For example, for the 14\,563 $z<0.75$ quasars in the \cite{Shen2011} Sloan Digital Sky Survey (SDSS) DR7 quasar catalog with reliable [\iona{O}{iii}] measurements, the 438 quasars with REW [\iona{O}{iii}]~$<3~\textup{\AA}$ have clearly large overall $\lambda_{\rm Edd}$ values (with a median value of 0.22) compared to the other REW [\iona{O}{iii}]~$>3~\textup{\AA}$ objects.

In this paper, we aim to test the X-ray photon-index selection of super-Eddington accreting quasars using the Chandra and XMM-Newton source catalogs.
We cross-verify the selected candidates using their $\lambda_{\rm Edd}$ parameter and the optical [\iona{O}{iii}] and \iona{Fe}{ii} emission strengths.
To this end, we obtained near-infrared (NIR) spectra of a pilot sample of nine $z\sim2$ quasars with $\Gamma$ values ranging from 2.0 to 2.6.
These rest-frame optical spectra allow more reliable SMBH mass estimates using the H$\beta$ emission line, and they offer [\iona{O}{iii}] REW and $R_{\rm Fe\ II}$ measurements.
The paper is organized as follows.
In Section~\ref{sec:data}, we describe the sample selection, NIR observations and data reduction, archival Chandra and XMM-Newton data reduction, and the comparison samples.
We present X-ray and optical spectral analyses in Section~\ref{sec:datreduc}.
In Section~\ref{sec:result}, 
we provide SMBH mass, Eddington ratio, and emission-line measurements, and compare the super-Eddington candidates to the comparison samples.
In Section~\ref{sec:dis}, we discuss the effectiveness of the \xray\ photon-index selection, and we report a good candidate with a strong [\iona{O}{iii}] outflow.
We summarize our results in Section~\ref{sec:sumf}. 
Throughout the paper, we adopt a cosmology with $H_0=67.4$~km~s$^{-1}$~Mpc$^{-1}$, $\Omega_{\rm M}=0.315$, $\Omega_{\Lambda}=0.685$ \citep{Planck2020}. Uncertainties are quoted at a $1\sigma$ confidence level, and upper limits are at a $3\sigma$ confidence level.

%%%%%%%%%%%%%%%%%%%%%%%%%%%%
\section{Sample Selection and Multiwavelength Observations} 
\label{sec:data}

%%%%%%%%%%%%%%%%%%%%%%%%%%%%
%sample selection
\subsection{Sample Selection and NIR Observations}
\label{subsec:sample}

Our targets were selected from the SDSS DR16 quasar catalog \citep{Wu2022}.
We stress that we aimed for a pilot sample as we were essentially limited by the available NIR observing time. Therefore, the selected sample is far from complete, and it is not necessarily representative either.
We chose \hbox{radio-quiet} quasars in the redshift range of $1.36 <z< 1.67$ or $2.1 <z< 2.48$. 
These redshift ranges were chosen to ensure that the H$\beta$ emission lines fall within the Palomar Hale 200-inch telescope (P200) wavelength coverage ($\approx$ \hbox{0.95--2.46}~$\rm \mu m$) and are not strongly affected by atmospheric absorption.
For economical observations, We selected bright quasars with $\textit{i}$-band magnitudes~$\leq 19$ and the Two Micron All Sky Survey (2MASS; \citealt{Skrutskie2006}) $\textit{H}$-band magnitudes $\leq 16.5$.
For quasars with no 2MASS $\textit{H}$-band detections, we estimated their $\textit{H}$-band magnitudes from their {Wide-field Infrared Survey Explorer} ({WISE}; \citealt{Wright2010}) $3.4~\mu m$ magnitudes adopting a typical $3.4~\mu m$--$\textit{H}$ color derived from other quasars with 2MASS $\textit{H}$-band and WISE $3.4~\mu m$ photometric measurements.
After applying these criteria, we obtained 3\,503 quasars.

%%% F1 mi vs. redshift pic
\begin{figure}
\centering \includegraphics[scale=0.33]{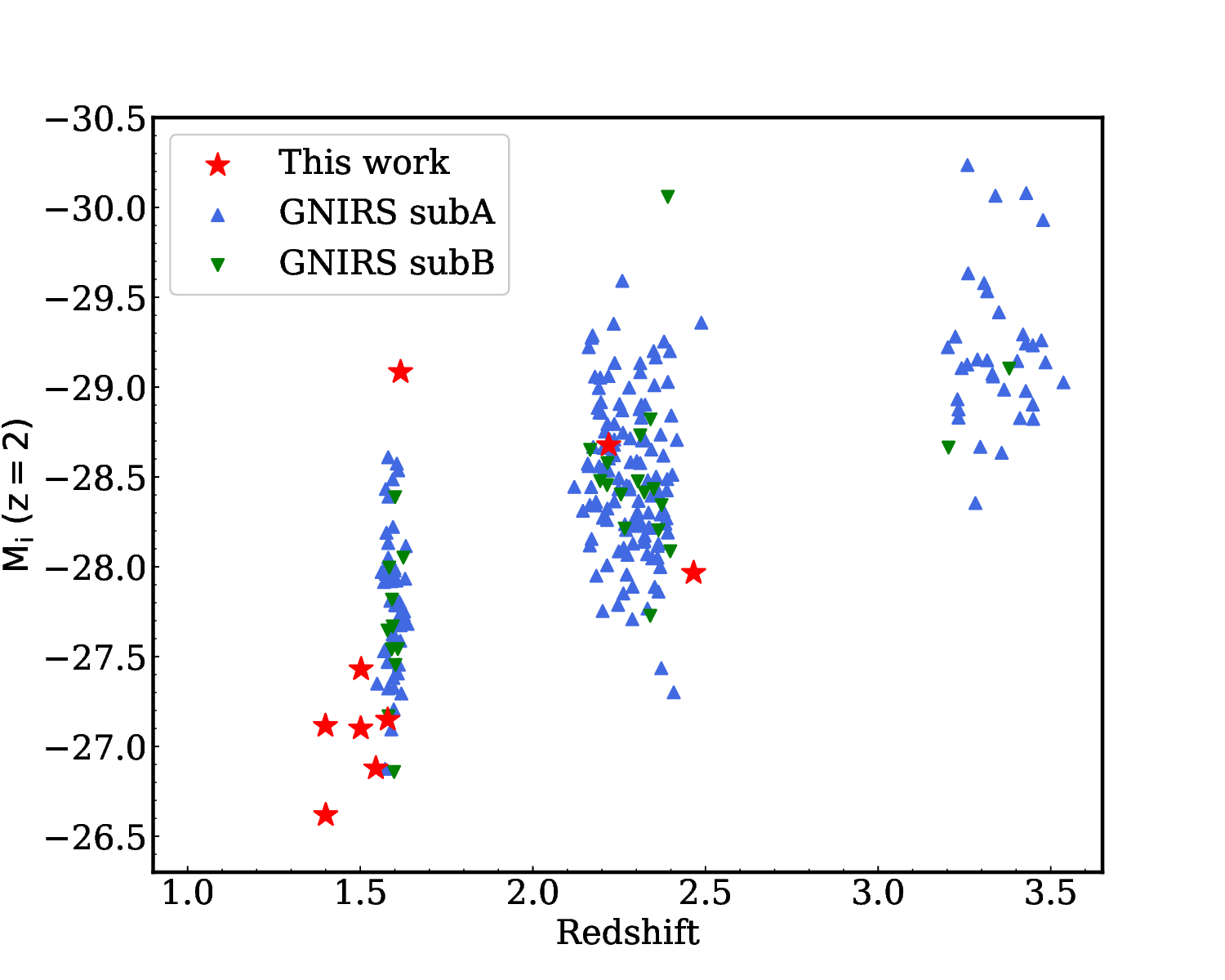}
\caption{Absolute $\textit{i}$-band magnitude vs.\ redshift for our nine targets, represented by the red stars.
The blue and green triangles represent the super-Eddington and sub-Eddington comparison samples drawn from \citet{Matthews2023}.
Our targets are on average less luminous than the GNIRS quasars.}
\label{fig-mi_z}
\end{figure}

We then searched for \xray\ properties of these quasars.
We cross-matched the 3\,503 quasars to the \chandra\ Source Catalog (CSC) Version 2.0 \citep{Evans2010} using a $3\arcsec$ matching radius, requiring more than 60 net counts in the \hbox{0.5--8~keV} band and \xray\ \hbox{power-law} photon indices ($\Gamma$) greater than 2.
There are 68 matches.
We then matched the 3\,503 quasars to the 4XMM-DR13 source catalog \citep{Webb2020} using a $5\arcsec$ matching radius, requiring more than 200 \hbox{0.2--12~keV} net counts\footnote{Different net count thresholds (60 versus 200) were used for the Chandra and XMM-Newton source catalogs due to the different effective areas and background levels of the two observatories. These thresholds were chosen empirically to yield $\Gamma$ measurements with comparable reliability levels.} from \xmm\ pn.
For these \xmm\ quasars, we estimated their \xray\ effective power-law photon indices ($\Gamma_{\rm eff}$) from the ratios of the \hbox{0.5--2~keV} and \hbox{2--12~keV} fluxes.
We selected 87 \xmm\ quasars with $\Gamma_{\rm eff}>2$. 
Excluding 21 objects in common, we obtained 134 quasars with large $\Gamma$ values from the \chandra\ and \xmm\ source catalogs.
We then filtered these sources by their visibility on our scheduled observational date, resulting in 39 candidates.
The pipeline-generated $\Gamma$ measurements are for the observed-frame \hbox{0.5--7.0~keV} or \hbox{0.5--12~keV} band, and they may also have large uncertainties (see Footnote~4).
Therefore, for each candidate, we processed its \chandra\ and/or \xmm\ observations and performed spectral fitting to obtain a more reliable hard \xray\ ($>2$~keV) $\Gamma$ measurement (see Sections~\ref{subsec:chandraobs}, \ref{subsec:xmmobs}, and \ref{subsec:xrayfit} below for analysis details).
Based on our $\Gamma$ measurements, we then identified 21 quasars that have $\Gamma>2$.

Our NIR observations were carried out with the Triple Spectrograph (TSpec) mounted on P200, which has spectral resolutions of $\approx$ \hbox{2500--2700} \citep{Herter2008}.
We obtained one night of P200/TSpec observations on 2022 February 11 via China's Telescope Access Program (TAP).\footnote{\url{https://tap.china-vo.org/}.}
Nine targets were observed that night with exposure times of either 40 minutes or 60 minutes.
We used a $1\arcsec \times 30\arcsec $ slit with standard ABBA nodding along the slit.
For flux calibration and telluric correction, we observed an A0V standard star either before or after each target observation.
The star was selected so that the airmass of its observation is comparable to that of the target.
The nine targets, out of the 21 $\Gamma>2$ quasars, were chosen based solely on their observing conditions (e.g., sky positions, airmasses), and they do not exhibit unusual X-ray properties compared to the other 12 quasars.
These nine quasars constitute the final sample of this study. 
The basic information of the sample objects and the P200 observations is presented in Table~\ref{tbl:obs}, and we show the absolute $i$-band magnitude versus redshift distributions for the nine quasars in Figure~\ref{fig-mi_z}.
The X-ray properties of the other 12 quasars are presented in Appendix~\ref{sec:appendix1}.

%%% T1 p200 and X-ray obs. log table
\begin{deluxetable*}{ccccccccccc}
\tabletypesize{\scriptsize}
\tablecolumns{11}
\tablecaption{Basic Object Properties and List of Observations for the Final Sample}
\tablehead{
\colhead{Source Name (SDSS J)}&
\colhead{$z_{\rm dr16}$}&
\colhead{$z_{\rm p200}$}&
\colhead{$i_{\rm psf}$}&
\colhead{$M_i$}&
\colhead{NIR P200/TSpec}&
\multicolumn{5}{c}{X-ray}\\
\cmidrule(r){7-11}
\colhead{Abbreviated name}&
\colhead{}&
\colhead{}&
\colhead{}&
\colhead{}&
\colhead{Exp. Time}&
\colhead{Observatory}&
\colhead{Obs. Date}&
\colhead{Observation}&
\colhead{Exp. Time}&
\colhead{Net Counts}\\
\colhead{}&
\colhead{}&
\colhead{}&
\colhead{(mag)}&
\colhead{(mag)}&
\colhead{(min)}&
\colhead{}&
\colhead{}&
\colhead{ID}&
\colhead{(ks)}&
\colhead{}
\\
\colhead{(1)}   &
\colhead{(2)}   &
\colhead{(3)}   &
\colhead{(4)}   &
\colhead{(5)}   &
\colhead{(6)}   &
\colhead{(7)}   &
\colhead{(8)}   &
\colhead{(9)}   &
\colhead{(10)}   &
\colhead{(11)}
}
\startdata
&                           &          &          &         & & Chandra  & 2010-09-27 & 12882          &  84.5     & 2415 \\
&                           &          &          &         & & Chandra  & 2015-07-08 & 17306          &  50.8     & 261 \\
$021830.59-045622.9$   & 1.400 & 1.402  & 17.78 & $-27.11$ & 60  & Chandra  & 2015-09-05 & 17311       &   48.8    & 160  \\
J0218&                      &          &          &         & & XMM-Newton  & 2000-07-31 & 0112370101  &     35.6  & 665/317/265  \\
&                           &          &          &         & & XMM-Newton  & 2000-08-02 & 0112371001  &     37.2  & 639/297/311 \\
&                           &          &          &         & & XMM-Newton  & 2017-02-09 & 0780452301  &     8.4   & 145/124/92 \\
\\
&                           &          &          &         & & XMM-Newton & 2001-07-03 & 0109520501  &     17.6   & 209/75/93 \\
&                           &          &          &         & & XMM-Newton & 2006-07-06 & 0404960501  &     7.7    & 51/35/47\\
$022354.80-044815.0$   & 2.452 & 2.466  & 18.42 & $-27.96$ & 60 &  XMM-Newton & 2016-08-13 & 0780450201  &  12.3   & 104/40/94\\
J0223&                      &          &          &         & & XMM-Newton & 2016-08-14 & 0780450301  &     10.9   & 122/--/43\\
&                           &          &          &         & & XMM-Newton & 2017-01-06 & 0780450501  &     9.6    & 160/94/18\\
&                           &          &          &         & & XMM-Newton & 2017-01-07 & 0780452201  &     7.1    & 76/--/36\\
&                           &          &          &         & & XMM-Newton & 2017-02-10 & 0780452501  &     7.7    & 45/45/43\\
\\
$080612.03+194853.6$   & 1.500 & 1.501  & 17.90 & $-27.10$ & 60 & Chandra & 2010-12-15 & 13015          &   19.7   & 87\\
J0806&                           &          &          &         & &  &  &   & \\
$083850.15+261105.4$   & 1.612 & 1.617  & 16.09 & $-29.08$ & 60 & XMM-Newton  & 2019-04-27 & 0821730301   &  16.0  & 928/447/430\\
J0838&                           &          &          &         & &  &  &   & \\
$102117.74+131545.9$   & 1.565 & 1.579  & 17.96 & $-27.14$ & 40 & XMM-Newton  & 2003-01-31 & 0146990101  &   15.8  & 264/114/120 \\
J1021&                           &          &          &         & &  &  &   & \\
&                           &          &          &         & & XMM-Newton  & 2003-05-05 & 0128531401  &    27.8   & 420/181/161 \\
$104401.13+212803.9$   & 1.501 & 1.502  & 17.58 & $-27.43$ & 40 & XMM-Newton  & 2003-05-28 & 0128531501  &  38.2   & 524/194/227\\
J1044&                           &          &          &    & & XMM-Newton  & 2003-12-12 & 0128531601  &    65.3   & 909/336/276\\
\\
&                           &          &          &         & & Chandra  & 2002-03-23 & 3253           &    8.8    & 29 \\
$111518.58+531452.7$   & 1.539 & 1.548  & 18.19 & $-26.87$ & 60 & Chandra  & 2004-06-22 & 5008           &   18.0  & 70\\
J1115&                           &          &          &    & & Chandra  & 2004-07-28 & 5350          &     6.9    & 22\\
&                           &          &          &         & & XMM-Newton  & 2003-04-25 & 0143650901  &    3.2    & 37/17/25 \\
\\
$121248.94+013727.1$   & 1.395 & 1.400  & 18.27 & $-26.61$ & 60 & XMM-Newton  & 2016-06-22 & 0760440101  &   19.5  & 261/88/94\\
J1212&                           &          &          &         & &  &  &   & \\
&                           &          &          &         & & Chandra  & 2002-09-25 & 3077$^{*}$           &  5.9 & 86\\
$142435.97+421030.4$   & 2.213 & 2.220  & 17.40 & $-28.67$ & 40 & XMM-Newton  & 2003-07-28 & 0148740801$^{*}$ & 1.7 & 65/62/73 \\
J1424&                           &          &          &    & & XMM-Newton  & 2003-12-18 & 0148742601$^{*}$  &  5.4 & 213/71/66 \\
\enddata
\tablecomments{
Column (1): Name and abbreviated name of the object, in order of increasing right ascension.
Column (2): Redshift derived from the SDSS DR16 quasar catalog \citep{Wu2022}.
Column (3): Redshift derived from the P200/TSpec observation.
Column (4): $i$-band PSF magnitude.
Column (5): Absolute $i$-band magnitude.
Columns (6): Exposure time for the P200/TSpec observation.
Columns (7), (8), and (9): \xray\ observatory, observation start data, and observation ID.
Column (10): Cleaned exposure time for the \xray\ observation. For \xmm\ observations, the pn exposure times are reported.
Column (11): 
Net source counts in the observed-frame 2/(1+$z$)--8~keV band for Chandra or the observed-frame 2/(1+$z$)--10~keV band for XMM-Newton pn/MOS1/MOS2 cameras; these are the counts in the spectra used for spectral fitting in Section~\ref{subsec:xrayfit}. For a few XMM-Newton observations, the target is not covered by all three cameras.
}
\tablenotetext{*}{These are targeted observations. The other observations are serendipitous coverage.}
\label{tbl:obs}
\end{deluxetable*}

\subsection{NIR Data Reduction}
\label{subsec:nirobs}

\begin{figure*}[h!]
\centering \includegraphics[scale=0.30]{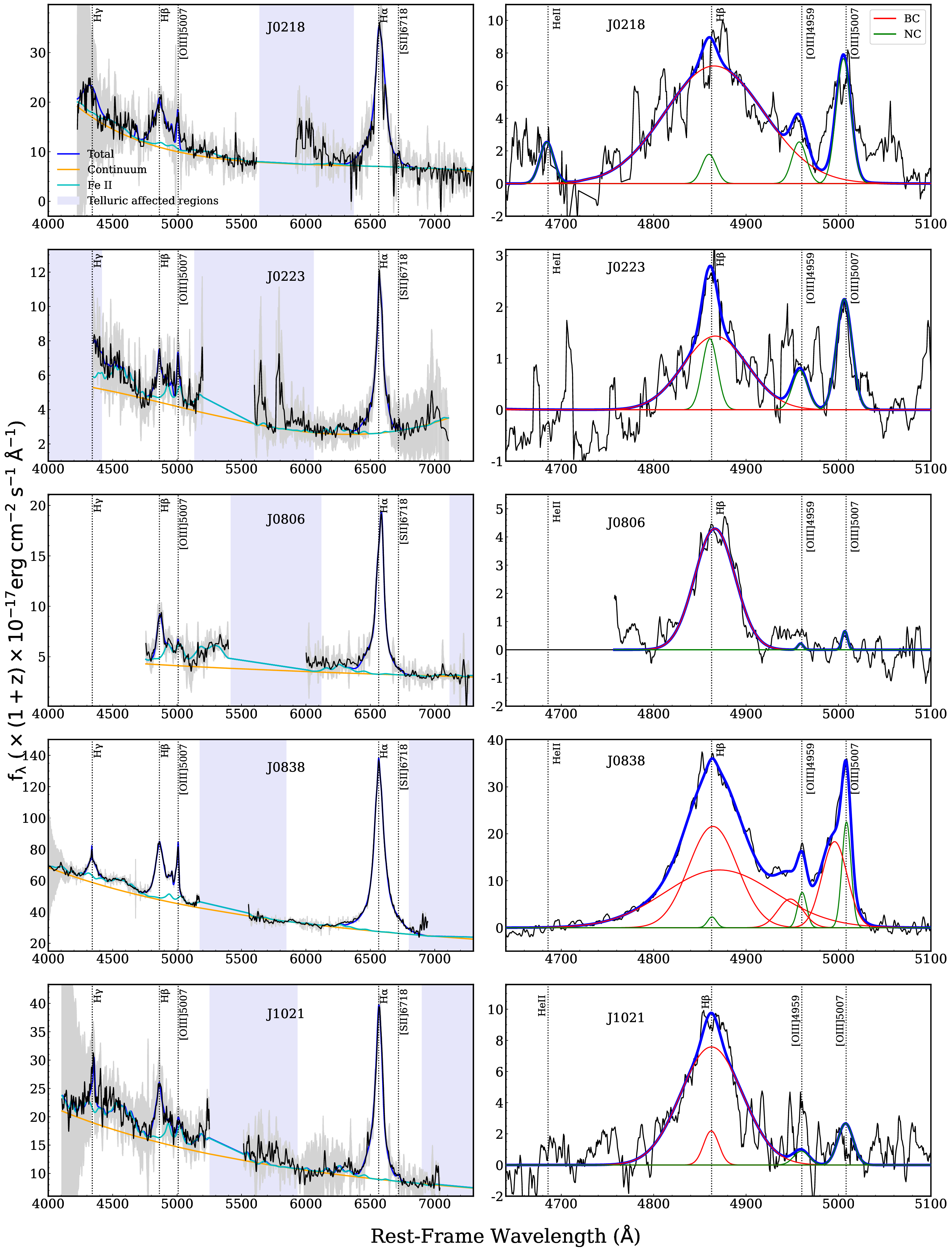}
\caption{
\hbox{Rest-frame} optical spectra for our nine targets with their best-fit models. 
The positions of major emission lines are marked for reference.
In the left panel, the \hbox{broad-band} spectra are represented by the black curves, smoothed using box convolutions varying from 5 to 10 pixels. 
Light gray shaded areas represent the spectral error bars (i.e., the uncertainty range of the spectral data); the best-fit models generally fall within these error regions.
The blue curves illustrate the \hbox{best-fit} continua $+$ \hbox{emission-line} models. 
The orange curves depict the \hbox{best-fit} continua, and cyan curves represent the \iona{Fe}{ii} pseudo continua.
The light blue shaded regions mark the portions of the spectra that are strongly affected by telluric absorption (observed-frame 13550--15300~$\textup{\AA}$ and 17800--21000~$\textup{\AA}$).
The right panel shows zoomed-in views of the spectra in the H$\rm \beta$ region after subtraction of the continuum and the \iona{Fe}{ii} pseudo continuum.
The H$\rm \beta$ and [\iona{O}{iii}] emission lines are fitted with broad (red curves) and/or narrow (green curves) Gaussian components.
}
\label{fig-spec1}
\end{figure*}

We reduced raw P200/TSpec observational data following the standard procedure detailed in Section~2.2 of \citet{Chen2024}.
Briefly, we constructed dark and flat-field files, performed wavelength calibration, extracted and stacked spectra, and then applied telluric absorption correction.
We evaluated the quality of the nine spectra obtained.
The average SNRs per pixel within the H$\beta$ region range from $\approx 4$ to $\approx 56$ with a median value around 7.
We compared our spectra with the available SDSS spectra by dividing them in the overlapping regions. 
The average ratios span from 0.45 to 1.53 with a median value of 1.24.
The flux differences likely arise from systematic uncertainties in the flux calibration procedure, SDSS flux calibration uncertainties, and quasar flux variability over year timescales.
The NIR spectra are displayed in Figure~\ref{fig-spec1}.

%%% F2 spectral and spectral fitting results pic
\renewcommand{\thefigure}{\arabic{figure} (Cont.)}
\addtocounter{figure}{-1}
\begin{figure*}
%\plotone{spec1.eps}
\centering \includegraphics[scale=0.30]{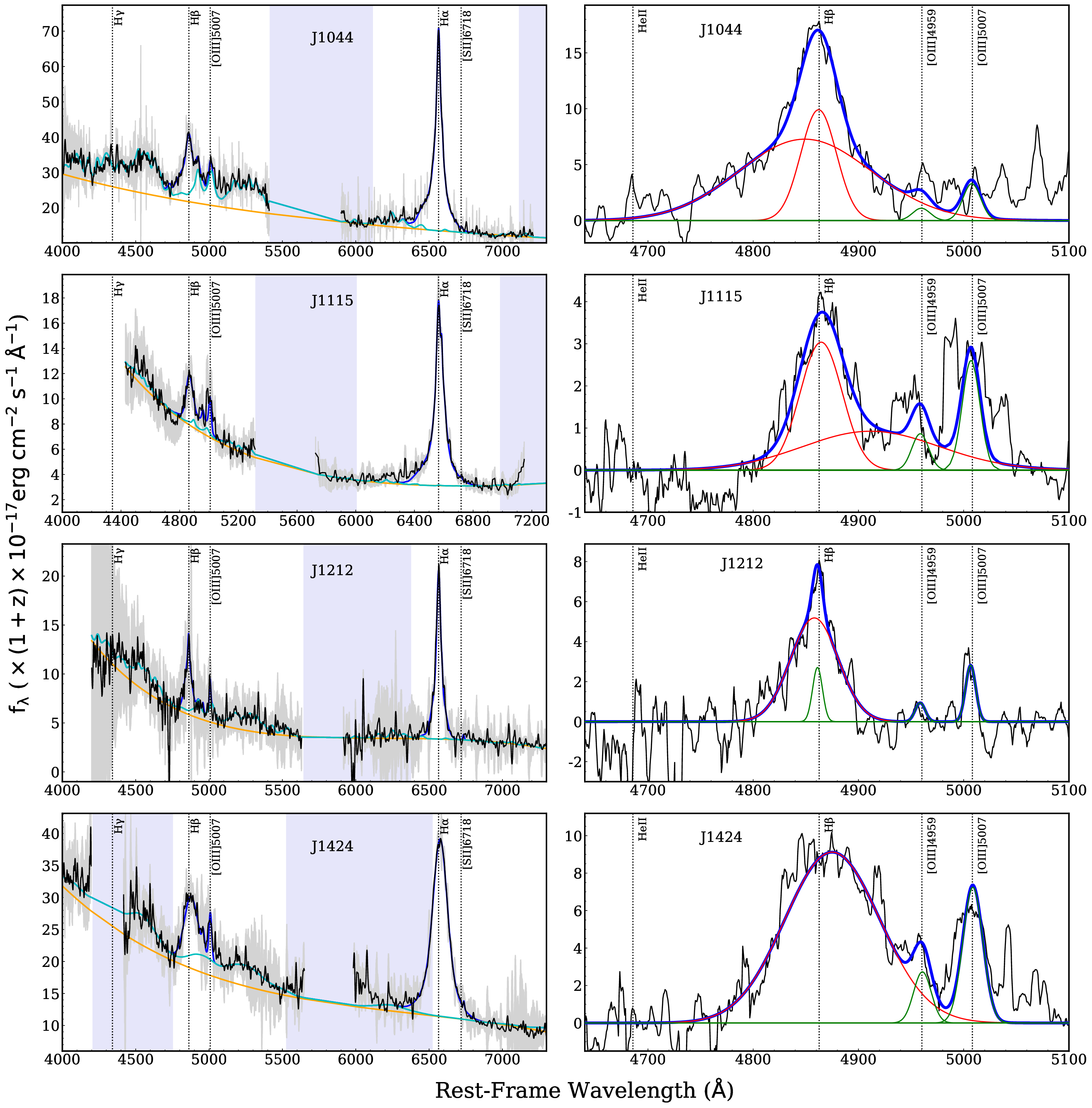}
\caption{}
\end{figure*}
\renewcommand{\thefigure}{\arabic{figure}}

\subsection{Archival Chandra Observations}
\label{subsec:chandraobs}

Four of the nine quasars have archival \chandra\ observations, listed in Table~\ref{tbl:obs}.
For \chandra\ data reduction, we used the Chandra Interactive Analysis of Observation (CIAO; v4.13) tool.
The {\sc chandra\_repro} script was used to generate new level 2 event files. 
Cleaned event files were created after filtering potential background flares with the {\sc deflare} script by an iterative 3$\sigma$ clipping algorithm.
We created an \hbox{0.5--8~keV} image from the cleaned event file by running the {\sc dmcopy} tool.
To search for \hbox{X-ray} sources in the \hbox{X-ray} images, we used the {\sc wavdetect} tool \citep{Freeman2002} with a false-positive  probability threshold of $10^{-6}$ and wavelet scales of 1, 1.414, 2, 2.828, 4, 5.656, and 8 pixels.
All these four quasars were detected within $0.19\arcsec$--$1.49\arcsec$ of the optical positions.
The source spectrum for each observation was extracted using {\sc specextract} with a circular source region centered on the \xray\ position.
For the targeted observation (ID 3077), we used an extraction radius of $2\arcsec$.
The other observations are serendipitous observations, and an extraction radius equal to 90\% the point spread function (PSF) size plus $3\arcsec$ was adopted for each of them.
We used the \hbox{{\sc psfsize\_src}} tool to determine the 90$\%$ PSF size at \hbox{1.5~keV}.
The background spectrum was extracted from an annular region centered on the \hbox{X-ray} position, with radii set at three times and five times the source extraction radius.
We verified that the background regions do not contain X-ray sources.
Specifically, for observation ID~17311 where the quasar is located near the chip edges, we adjusted the location and size of the background region.
Two of these quasars have multiple archival observations.
The maximum flux variation amplitudes between these observations appear small ($\approx 33\%-42\%$), and the spectral shapes do not appear to vary significantly either (consistent power-law $\Gamma$ values considering the uncertainties). Therefore, we employed the \hbox{{\sc combine\_spectra}} script to combine these multi-epoch observations for these targets.
The combined spectra are utilized for the spectral fitting in Section~\ref{subsec:xrayfit} below.
The Chandra spectrum for J0806 is shown in Figure~\ref{fig-chandra} as an example.

%%%%%%%%%%%%%%%%%%%%%%%%%%%%
\subsection{Archival XMM-Newton Observations}
\label{subsec:xmmobs}

Eight of the nine quasars have \xmm\ observations, listed in Table~\ref{tbl:obs}.
For the \xmm\ observations, we used the data from the EPIC pn \citep{Struder2001} and EPIC MOS (MOS1 and MOS2; \citealt{Turner2001}) cameras.
We used the Science Analysis System (SAS; v19.1.0) to process the data following the standard procedure in the SAS Data Analysis Threads.\footnote{http://www.cosmos.esa.int/web/xmm-newton/sas-threads.}
We obtained calibrated and concatenated event files using the task {\sc epproc} for the pn camera and {\sc emproc} for the MOS cameras.
Thresholds of \hbox{0.4 ${\rm cts}~{\rm s}^{-1}$} (pn) and \hbox{0.35 ${\rm cts}~{\rm s}^{-1}$} (MOS) were adopted to filter background flares.
We created \hbox{good-time-interval} files using the {\sc tabgtigen} task, and we generated cleaned event files using the {\sc evselect} tool.
For each detector, we used the task {\sc evselect} to extract the source and background spectra; 
the source region is a \hbox{30\arcsec-radius} circular region centered on the optical position of the quasar,
and the background region is a \hbox{50\arcsec-radius} circular \hbox{source-free} region on the same CCD chip.
We determined the source significance by calculating the binomial no-source probability ($P_{\rm B}$; e.g., \citealt{Luo2015}), which is defined as
\begin{equation}
P_{\rm B}(X\ge S)=\sum_{X=S}^{N}\frac{N!}{X!(N-X)!}p^X(1-p)^{N-X}~.
\end{equation}
In this expression, $S$ and $B$ are source and background counts in the \hbox{$2/(1+z)$--10 keV} band extracted from the same source and background regions as in the spectral extraction,
$N$ = $S+B$, and $p = 1/(1 +BACKSCAL$), 
where $BACKSCAL$ is the ratio between the areas of the background and source regions.
A smaller $P_{\rm B}$ value indicates a more significant signal. 
We considered the source detected if the measured $P_{\rm B}$ value is smaller than 0.01 (corresponding to a $>2.6 \sigma$ significance level).
We used only the \xmm\ observations where the target is detected; three pn and five MOS observations were thus excluded. 
Four quasars have multi-epoch archival data, showing only mild flux variability (maximum flux variability amplitudes between 6\% and 65\% across different observations)
and minimal spectral shape variability (consistent power-law $\Gamma$ values).
Thus we used the {\sc epicspeccombine} script to combine the multi-epoch spectra for each camera.
The combined pn and MOS spectra of J0223 are shown in Figure~\ref{fig-xmm} as an example.

Five of our sources have observations from the \xmm\ Optical Monitor (OM; \citealt{Mason2001}).
For each source, one to three filters were used from the five OM optical/UV filters: UVW2, UVM2, UVW1, U, and B, with effective wavelengths of 2120~\AA, 2310~\AA, 2910~\AA, 3440~\AA, and 4500~\AA, respectively.
We processed the OM data and generated 12 exposures using the pipeline task {\sc omichain}.
The source flux and magnitude measurements for each exposure were extracted from the SWSRLI files.
For J1424, the only target observed with two exposures using the UVW1 filter, we adopted the mean magnitude for this filter.

%%% F3 chandra spectral fitting results pic
\begin{figure}
\centering \includegraphics[scale=0.33]{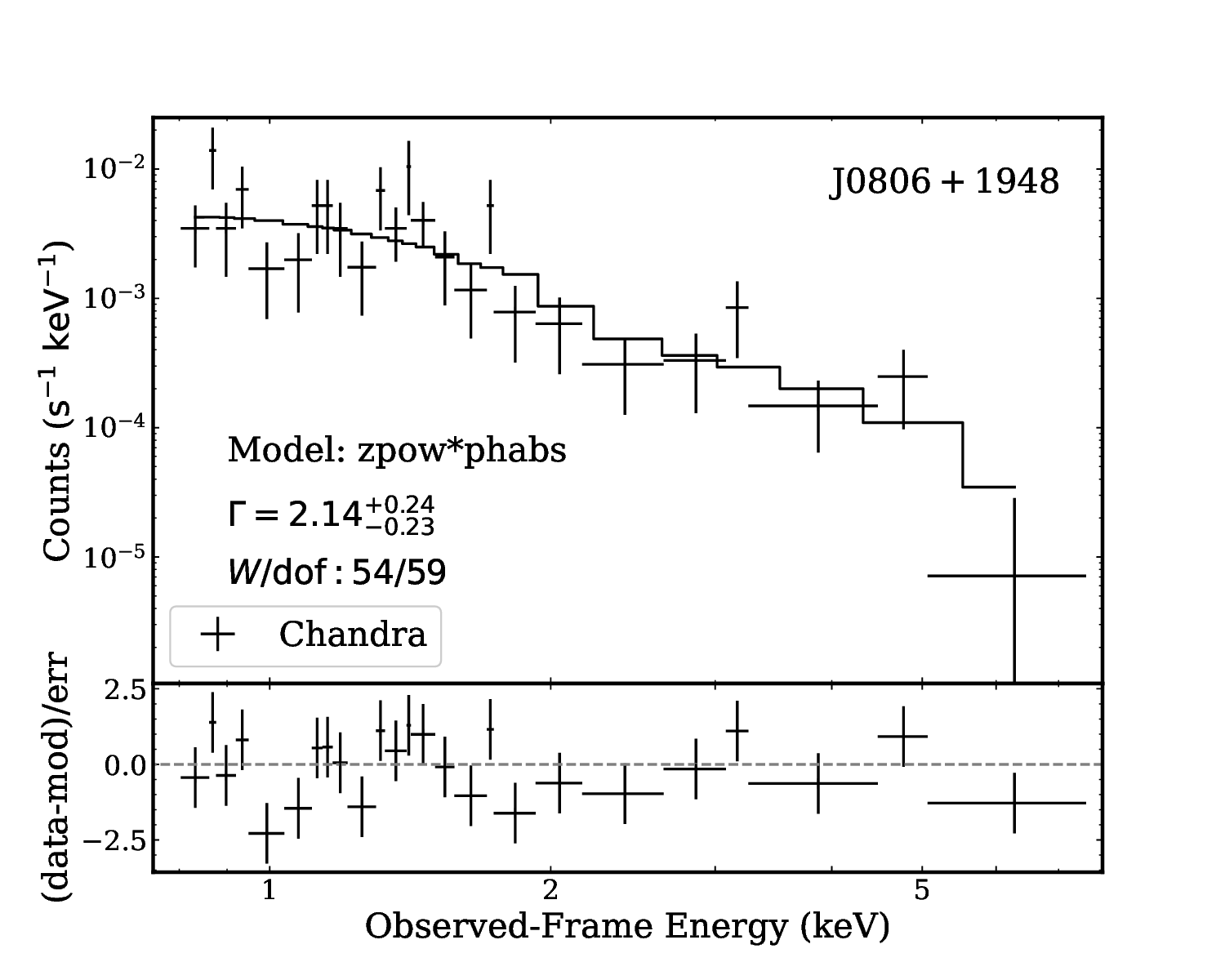}
\caption{The \chandra\ spectrum for J0806 overlaid with the best-fit simple power-law model.
The bottom panel displays the residuals between the spectral data and the best-fit model, normalized by the errors.}
\label{fig-chandra}
\end{figure}

%%% F4 xmm-newton spectral fitting results pic
\begin{figure}
\centering \includegraphics[scale=0.33]{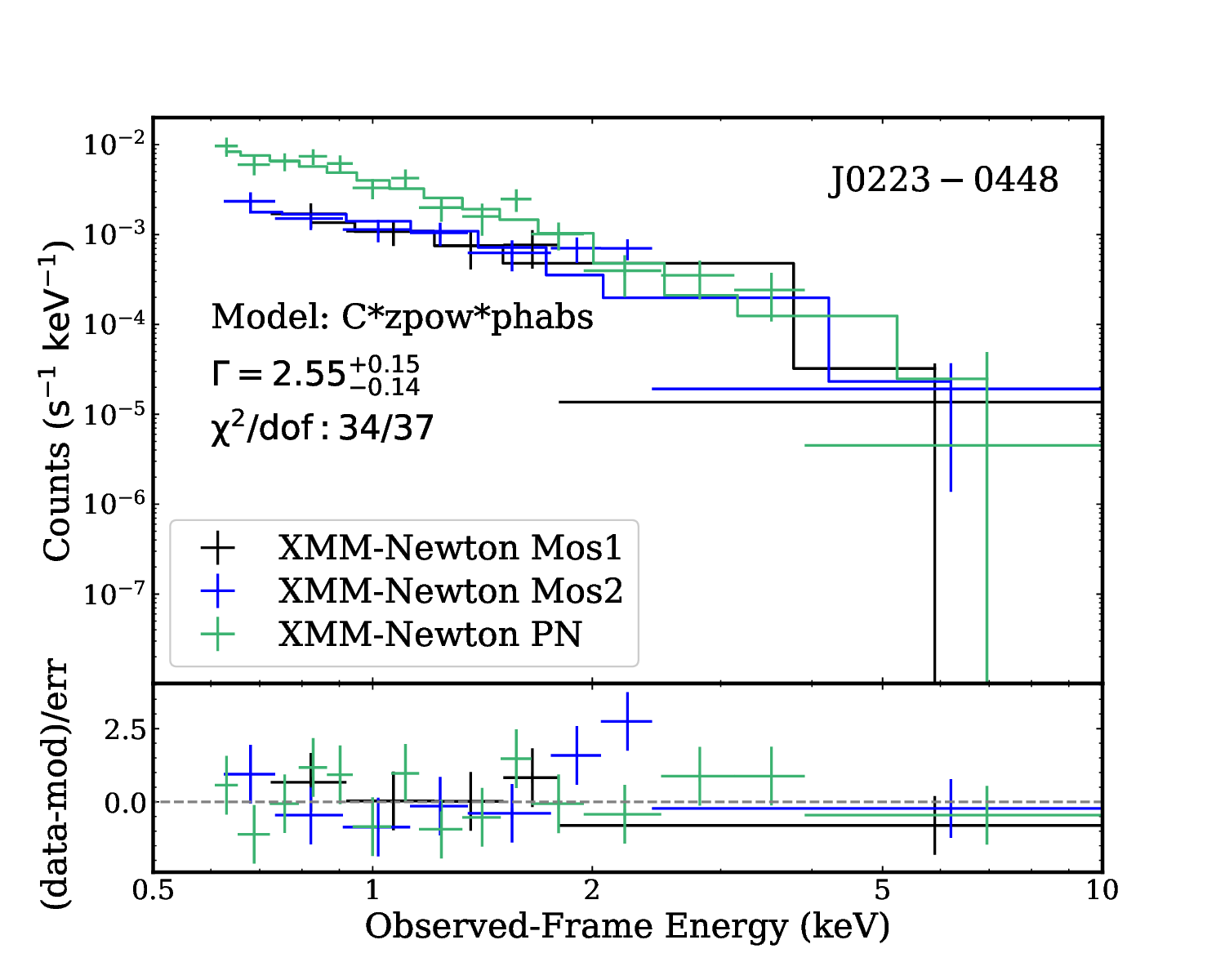}
\caption{Similar to Figure~\ref{fig-chandra}, displaying the combined \xmm\ pn and Mos spectra for J0223
overlaid with the best-fit simple power-law model.
The \xmm\ pn, MOS1, and MOS2 spectra and their corresponding best-fit models are shown in green, black, and blue, respectively.
The bottom panel displays the residuals between the spectral data and the best-fit model, normalized by the errors.}
\label{fig-xmm}
\end{figure}

%%% T2 X-ray prop. table
\begin{deluxetable*}{lccccccccc}
\tablewidth{0pt}
\tablecaption{Best-fit Parameters from Joint Spectral Fitting of the Multi-epoch X-ray Spectra}
\tablehead{
\colhead{Source Name}&
\colhead{$N_{\rm H, Gal}$}&%2
\colhead{$\Gamma$}&
\multicolumn{3}{c}{C}&
\colhead{$\rm \chi^2/dof$}&
\colhead{$P_{\rm null}$}&
\colhead{$f_{\rm 2~keV}$}& 
\colhead{$L_{\rm 2-10~keV}$}
\\  
\cmidrule(r){4-6}
\colhead{(SDSS J)}&
\colhead{$\rm (10^{20}~{\rm cm}^{-2})$} &%2
\colhead{} &
\colhead{MOS1}  &
\colhead{MOS2}  &
\colhead{\chandra}  &
\colhead{($W$/dof)}  &
\colhead{}& 
\colhead{($\rm 10^{-31}~erg~cm^{-2}~s^{-1}~Hz^{-1}$)} &
\colhead{($\rm 10^{44}~{\rm erg}~~{\rm s}^{-1}$)}  
\\
\colhead{(1)}   &
\colhead{(2)}   &
\colhead{(3)}   &
\colhead{}   &
\colhead{(4)}   &
\colhead{}   &
\colhead{(5)}   &
\colhead{(6)} &
\colhead{(7)}   &
\colhead{(8)} 
} 
\startdata
$021830.59-045622.9$ & 1.97  & $2.09_{-0.05}^{+0.05}$  &1.18&1.03&1.33& 202/187 & 0.21 & $1.33_{-0.05}^{+0.06}$& $5.01_{-0.21}^{+0.22}$  \\
$022354.80-044815.0$ & 2.16  & $2.55_{-0.14}^{+0.15}$  &0.85&1.32&--& 34/37 & 0.59     & $1.08_{-0.13}^{+0.15}$& $7.99_{-0.48}^{+0.81}$  \\
$080612.03+194853.6$ & 2.94  & $2.14_{-0.23}^{+0.24}$  &--&--&--& (54/59) & --         & $0.83_{-0.11}^{+0.19}$& $3.40_{-0.27}^{+0.61}$   \\ %bin1 cstat
$083850.15+261105.4$ & 3.18  & $2.02_{-0.06}^{+0.06}$  &1.07&1.15&--& 66/58 & 0.21     & $3.44_{-0.22}^{+0.19}$& $17.86_{-0.70}^{+0.87}$  \\
$102117.74+131545.9$ & 3.95  & $2.20_{-0.13}^{+0.14}$  &0.91&1.12&--& (356/382) & --   & $2.19_{-0.21}^{+0.22}$& $9.49_{-0.92}^{+0.73}$  \\ %bin1 cstat
$104401.13+212803.9$ & 1.64  & $2.26_{-0.07}^{+0.07}$  &1.00&0.95&--& 115/106 & 0.27   & $1.73_{-0.09}^{+0.06}$& $6.49_{-0.26}^{+0.20}$   \\
$111518.58+531452.7$ & 0.76  & $2.29_{-0.21}^{+0.22}$  &--&--&1.34& (100/110) & --     & $0.81_{-0.26}^{+0.28}$& $3.15_{-1.10}^{+0.57}$   \\ %bin1 cstat
$121248.94+013727.1$ & 2.17  & $2.08_{-0.18}^{+0.19}$  &1.12&1.32&--& (358/338) & --   & $1.26_{-0.14}^{+0.17}$& $4.75_{-0.59}^{+0.28}$  \\ %bin1 cstat
$142435.97+421030.4$ & 0.81  & $2.02_{-0.09}^{+0.09}$  &0.89&0.82&0.99& (423/467) & -- & $2.50_{-0.27}^{+0.30}$&  $23.14_{-1.77}^{+1.64}$ \\ %bin1 cstat
\enddata
\tablecomments{Column (1): Name of the object. Column (2): Galactic neutral hydrogen column density. Column (3): Hard \xray\ photon index. Column (4): Multiplicative constant factors for spectra from different instruments. Column (5): $\chi^2$ statistic value over degree of freedom or $W$ statistic value over degree of freedom in parentheses. Column (6): Null-hypothesis probability for the $\chi^2$ statistic. Column (7):  Flux density at \hbox{rest-frame} \hbox{2 keV}, corrected for the Galactic absorption. Column (8): Luminosity in the \hbox{rest-frame} \hbox{2--10 keV} band, corrected for the Galactic absorption.}
\label{tab:x_ray}
\end{deluxetable*}

%%%%%%%%%%%%%%%%%%%%%%%%%%%%
\subsection{Comparison Samples}
\label{subsec:comp}

To assess whether our quasars exhibit distinct rest-frame optical emission-line features compared to typical quasars or sub-Eddington accreting quasars at similar redshifts, we constructed comparison samples of quasars drawn from the Gemini Near Infrared Spectrograph-Distant Quasar Survey (GNIRS-DQS; \citealt{Matthews2021,Matthews2023}),
a homogeneous flux-limited NIR spectroscopic survey of 260 high-redshift quasars.
Following Section~2.4 of \citet{Chen2024}, we excluded significantly \hbox{radio-loud} ($R>100$)\footnote{The \hbox{radio-loudness} parameter ($R$) is defined as $R=f_{5~{\rm GHz}}/f_{\rm 4400~{\textup{\AA}}}$ \citep{Kellermann1989}, where $f_{5~{\rm GHz}}$ and $f_{\rm 4400~{\textup{\AA}}}$ are the flux densities at 5~GHz and $\rm 4400~{\textup{\AA}}$, respectively.}
quasars from the sample, as the optical continua of these quasars might have potential contamination from \hbox{non-thermal} jet radiation.
The remaining 240 quasars constitute our full comparison sample (the GNIRS sample).
We show the absolute $i$-band magnitude versus redshift distributions for the comparison samples in Figure~\ref{fig-mi_z}.
Overall, our nine quasars are less luminous than the GNIRS quasars.

We note that the $\approx 0.4$--0.5~dex differences in the median luminosities and SMBH masses (see Section~\ref{subsec:bhmledd} below) of our sample and the GNIRS sample should not significantly influence the comparison results in Section~\ref{sec:result}, as the optical emission-line properties of quasars do not generally display strong evolution within such a luminosity range.
There is a negative correlation between the [\iona{O}{iii}] REW and quasar continuum luminosity, i.e., more luminous quasars tend to exhibit weaker [\iona{O}{iii}]~$\lambda 5007$ lines \citep[e.g.,][]{Sulentic2004,Netzer2006,Stern2012,Shen2016,Coatman2019}.
This is considered an analogy to the Baldwin effect originally found for the \iona{C}{iv}~$\lambda 1549$ line \citep{Baldwin1977}.
The expected [\iona{O}{iii}] REW difference from the luminosity difference of our sample and the GNIRS sample is negligible (e.g., Figure 6 of \citealt{Coatman2019}), and even if present, this Baldwin effect would only enhance the contrast found in Section~\ref{sec:result} that our less luminous quasars show weaker [\iona{O}{iii}] lines.

For the GNIRS sample, measurements of the H$\beta$, [\iona{O}{iii}], and \iona{Fe}{ii}  emission lines in \citep{Matthews2023} are adopted in our following comparisons.
An [\iona{O}{iii}] REW upper limit of $5~\textup{\AA}$ was adopted for GNIRS objects with [\iona{O}{iii}] REW below $1~\textup{\AA}$ \citep{Chen2024}.
For the one quasar (J1424) in common of our sample and the GNIRS sample, we verified that the measurements of the emission-line properties are consistent within the errors.

We caution that although the GNIRS sample should be representative of $z\sim 2$--3 high-luminosity quasars, it might actually contain a large fraction of super-Eddington accreting quasars, 
as the high luminosities could be a consequence of high accretion rates and super-Eddington accreting quasars could also prevail in the high-redshift universe \citep[e.g.,][]{Netzer2007,Shen2012}.
In fact, 211 of 240 quasars have $\lambda_{\rm Edd,FeII}>0.3$ (see Section~\ref{subsec:bhmledd} below for $\lambda_{\rm Edd,FeII}$ estimates).
Therefore, we extracted two subsamples from the GNIRS sample, separate at $\lambda_{\rm Edd,FeII}=0.3$; the 211 quasars with $\lambda_{\rm Edd,FeII}\geq0.3$ constitute the GNIRS subA sample and the 29 quasars with $\lambda_{\rm Edd,FeII}<0.3$ constitute the GNIRS subB sample.
As introduced in Section~\ref{sec:intro}, there is no clear $\lambda_{\rm Edd}$ threshold for identifying super-Eddington accreting AGNs, and in this study we adopt 0.3 as the threshold to divide the GNIRS sample in to the super-Eddington and sub-Eddington subsamples.
The $\lambda_{\rm Edd,FeII}$ parameter was computed using the \cite{DU2019} updated $R$--$L$ relation (see Section~\ref{subsec:bhmledd} below).
We verified that using the traditional $R$--$L$ relation would result in 157 quasars in the GNIRS subA sample and 83 quasars in the GNIRS subB sample, but the qualitative comparison results remain the same.
In our following analysis, we compare our source properties to those of the GNIRS, GNIRS subA, and GNIRS subB samples.

%%% emission line prop. table
\begin{deluxetable*}{lccccccc}
\tablewidth{0pt}
\tablecaption{Emission-Line Properties}
\tablehead{
\colhead{Source Name}&
\colhead{$\rm H \beta$~REW}&
\colhead{[\iona{O}{iii}]~$\lambda 5007$~REW}&
\colhead{\iona{Fe}{ii}~REW}&
\colhead{$\rm H \beta$~FWHM}&
\colhead{$R_{\rm Fe\ II}$}&
\colhead{\iona{C}{iv}~REW}& \\
\colhead{(SDSS J)}&
\colhead{($\rm \textup{\AA}$)}&
\colhead{($\rm \textup{\AA}$)}&
\colhead{($\rm \textup{\AA}$)}&
\colhead{($\rm km~s^{-1}$)}&
\colhead{}&
\colhead{($\rm \textup{\AA}$)}& \\
\colhead{(1)}   &
\colhead{(2)}   &
\colhead{(3)}   &
\colhead{(4)}   &
\colhead{(5)}   &
\colhead{(6)}   &
\colhead{(7)}
}
\startdata
$021830.59-045622.9$  & $122.3\pm9.2$& $10.0\pm1.9$&  $23\pm8$&  $5620\pm610$&  $0.2\pm0.1$ & $39.0\pm1.2$ \\
$022354.80-044815.0$  & $37.4\pm3.5$&  $9.3\pm2.1$&  $56\pm27$&  $3320\pm750$&  $1.5\pm0.4$ & $24.0\pm0.6$ \\
$080612.03+194853.6$  & $55.6\pm3.1$&  $1.2\pm0.3$&  $138\pm13$& $3160\pm130$&  $2.5\pm0.4$ & $21.6\pm2.0$ \\
$083850.15+261105.4$  & $68.5\pm0.7$&  $6.1\pm0.8$&  $25\pm2$&   $4870\pm100$&  $0.4\pm0.1$ & $31.4\pm2.1$ \\
$102117.74+131545.9$  & $59.9\pm7.9$&  $2.7\pm1.5$&  $52\pm16$&  $3720\pm260$&  $0.9\pm0.2$ & $11.7\pm0.4$ \\
$104401.13+212803.9$  & $93.4\pm17.0$& $0.7\pm2.3$&  $92\pm12$&  $3910\pm1640$& $1.0\pm0.3$ & $12.1\pm0.7$ \\
$111518.58+531452.7$  & $39.1\pm4.2$&  $8.5\pm1.9$&  $11\pm6$&   $3280\pm540$&  $0.3\pm0.2$ & $35.7\pm2.6$ \\
$121248.94+013727.1$  & $55.9\pm4.4$&  $7.8\pm2.1$&  $52\pm12$&  $3570\pm400$&  $0.9\pm0.1$ & -- \\
$142435.97+421030.4$  & $61.6\pm2.4$&  $7.2\pm0.7$&  $41\pm3$&   $6070\pm290$&  $0.7\pm0.1$ & $29.2\pm1.0$ \\
\enddata
\tablecomments{Column (1): Name of the object.  Columns (2), (3), and (4): REWs for broad H$\beta$, narrow [\iona{O}{iii}]~$\lambda 5007$, and optical \iona{Fe}{ii}. Columns (5): FWHM for broad H$\beta$. Column (6): The relative strength of the optical \iona{Fe}{ii} emission to the broad H$\beta$ emission, $R_{\rm Fe\ II}$. Column (7): \iona{C}{iv} REW, adopted from \citet{Wu2022}. There is no \iona{C}{iv} coverage for J1212.}
\label{tab:elp}
\end{deluxetable*}

%%%%%%%%%%%%%%%%%%%%%%%%%%%%
\section{Data Analysis}
\label{sec:datreduc}

\subsection{X-ray Spectral Fitting}
\label{subsec:xrayfit}

Our X-ray spectral fitting was conducted utilizing the HEASoft (v6.33.1) tool XSPEC (v.12.14.0b; \citealt{Arnaud1996}).~% check: define heasoft
We jointly fitted the \chandra\ and \xmm\ spectra (including pn, MOS1, and MOS2) when present.
The energy ranges considered for the spectral fitting were \hbox{$2/(1+z)$--8 keV} for the \chandra\ spectra and \hbox{$2/(1+z)$--10 keV} for the \xmm\ spectra.
For J0806, J1021, J1115, J1212 and J1424, we used the $W$ statistic in XSPEC and the spectra were grouped with at least one count per bin.
For the other sources, the $\chi^{2}$ statistic was applied and the spectra were grouped with at least 25 counts per bin.

A simple \hbox{power-law} model adjusted by Galactic absorption ({\sc zpowerlw*phabs}) was adopted for all the spectra.
For each source, the Galactic neutral hydrogen column density $N_{\rm H}$ was determined with the HEASoft {\sc nh} tool (v.3; \citealt{HI4PI2016}).
To account for possible calibration uncertainties among different instruments and mild flux variations between the \chandra\ and \xmm\ observation epochs, we included a multiplicative constant factor (C) in the fit.
Only J0806 does not need these cross-normalization factors as it has a single Chandra spectrum. 
We fixed the C value for the pn spectra at 1.0,
the resulting C values range from 0.81 to 1.34 for the MOS and Chandra spectra.
The best-fit results are presented in Table~\ref{tab:x_ray}, and the best-fit models for J0806 and J0223 are displayed in Figure~\ref{fig-chandra} and Figure~\ref{fig-xmm}, respectively.
The simple \hbox{power-law} model provides acceptable descriptions of the overall \chandra\ and \xmm\ spectra (as indicated by the small $\chi^2$/dof or $W$/dof values).
The best-fit $\Gamma$ values for our quasars range from 2.0 to 2.6, as per our sample construction (Section~\ref{subsec:sample}).
Based on the \hbox{best-fit} models, we derived \hbox{rest-frame} \hbox{2 keV} flux densities and \hbox{2--10} keV luminosities, corrected for the Galactic absorption.

\subsection{NIR Spectral Analysis}
\label{subsec:nirfit}

The P200/TSpec NIR spectra were fitted following Section~2.3 of \citet{Chen2024}.
We utilized the PyQSOFit package (v1.1)\footnote{https://github.com/legolason/PyQSOFit.} to fit the continuum and emission lines \citep[e.g.,][]{Shen2011,pyqsofit,Guo2019,Wu2022}. 
Galactic extinction corrections were firstly applied using the $E(B-V)$ values from \citet{Schlegel1998}, following the de-reddening approach in \citet{Cardelli1989} and \citet{O'Donnell1994}.
For each spectrum, we fit the continuum with a power-law component, a polynomial component, and the \citet{Boroson1992} optical \iona{Fe}{ii} emission template.
The polynomial component is used to account for spectral complexity such as reddening, and it is only needed for five sources.
The fitting was performed using line-free windows around the H$\alpha$ and H$\beta$ emission lines (i.e., 4435--4630~$\textup{\AA}$, 5100--5535~$\textup{\AA}$, 6000--6250~$\textup{\AA}$, and 6800--7000~$\textup{\AA}$).
Given the best-fit continuum, we derived the monochromatic luminosity at 5100~$\textup{\AA}$ ($L_{\rm 5100}$) from the power-law component $+$ the polynomial component (if present), and we measured the REW of the \iona{Fe}{ii} pseudo continuum between 4434~$\textup{\AA}$ and 4684~$\textup{\AA}$ against the underlying continuum.

After subtracting the continuum, we used two groups of Gaussian profiles to fit the emission lines.
In the H$\beta$ region, we used four Gaussians for the [\iona{O}{iii}] doublet: two narrow (full width at half maximum FWHM $\rm <1200~km~s^{-1}$) and two broad (FWHM $\rm \geq 1200~km~s^{-1}$).
Only one object, J0838, appears to have broad [\iona{O}{iii}] components.
The broad components have significant blueshifts and might indicate large-scale ionized outflows in this quasar (see Section~\ref{subsec:outflow} below), and thus we did not include the broad [\iona{O}{iii}]~$\lambda 5007$ component in its [\iona{O}{iii}] REW measurement.
For the H$\beta$ line, we used one Gaussian for the narrow component and up to two Gaussians for the broad component.
For the H$\alpha$ complex, we used up to four narrow Gaussians to model the [\iona{N}{ii}] and [\iona{S}{ii}] doublets, one Gaussian for the narrow H$\alpha$ line, and up to three Gaussians for the broad H$\alpha$ line.
We fixed the flux ratios of the narrow [\iona{O}{iii}] doublet and [\iona{N}{ii}] doublet components to their theoretical value of 3.0 (i.e., $f_{5007}/f_{4959} = 3$ and $f_{6584}/f_{6548} = 3$).
Addtionally, the velocity offsets and line widths of the narrow components in each group are tied.
Multiple Gaussian profiles are widely used for quasar spectral fitting, since a single Gaussian generally cannot reproduce the broad emission-line shapes \citep[e.g.,][]{Shen2011,Plotkin2015,DU2019,Guo2019,Matthews2021,Matthews2023,Chen2024}. 
The best-fit individual components are not necessarily physically meaningful, and line parameters (e.g., REW, FWHM, flux) are measured from the combined profile.
From the combined broad or narrow component for each line, we calculated the line REW and FWHM.

The spectra and the best-fit model are displayed in Figure~\ref{fig-spec1}, 
Our NIR spectra have limited SNRs.
Large residuals are sometimes present in the telluric absorption regions (light blue shaded regions in Figure~\ref{fig-spec1}), likely caused by uncertainties in telluric absorption correction.
Additional residuals mainly occur redward of the H$\beta$ line, probably related to uncertainties in the \iona{Fe}{ii} pseudo-continuum modeling.
We tested other optical \iona{Fe}{ii} templates, including those from \citet{Cetty2004}, \citet{Kova2010}, and the more recent work by \citet{Pandey2025}.
However, none of these templates significantly improves the fits.
We note that our quasars generally exhibit strong \iona{Fe}{ii} emission, and the residuals do not contribute a significant flux compared to the current best-fit \iona{Fe}{ii} pseudo continuum.
Therefore, the $R_{\rm Fe\ II}$ measurements are not significantly affected.
The emission-line measurements are summarized in Table~\ref{tab:elp}.
We employed a Monte Carlo approach to assess the statistical uncertainties of the parameters we measured. 
We generated 200 mock spectra by adding Gaussian noise to each spectrum using the flux density errors (gray shaded areas in Figure~\ref{fig-spec1}), fitted the mock spectra with the same routine, and derived the scatter of each parameter.

We examined the redshifts of our sources using the P200/TSpec spectra.
Their redshifts from the SDSS DR16 quasar catalog \citep{Wu2022} are listed in Table~\ref{tbl:obs}.
These redshifts were derived from the broad UV emission lines (i.e., \iona{C}{iv} and \iona{Mg}{ii}).
We obtained redshifts of our sources from the peak positions of the narrow [\iona{O}{iii}] emission-line profiles adopting an iterative procedure.
The \citet{Wu2022} redshift was used as the initial input in PyQSOFit. 
We then measured a new redshift from the average velocity offset of the best-fit [\iona{O}{iii}] narrow component. 
The new redshift was used in the next iteration. 
If the redshift difference is smaller than 0.001, we considered the value converged and stopped the iteration. 
The resulting redshifts are listed in Table~\ref{tbl:obs}, and they differ slightly (within the range of $-0.008$--0.004 with an average value of $-0.001$) from the \citet{Wu2022} redshifts. 
We adopt the P200/TSpec redshifts in the following analyses.

%%% F7 mbh compare pic
\begin{figure}
\centering \includegraphics[width=\linewidth,keepaspectratio]{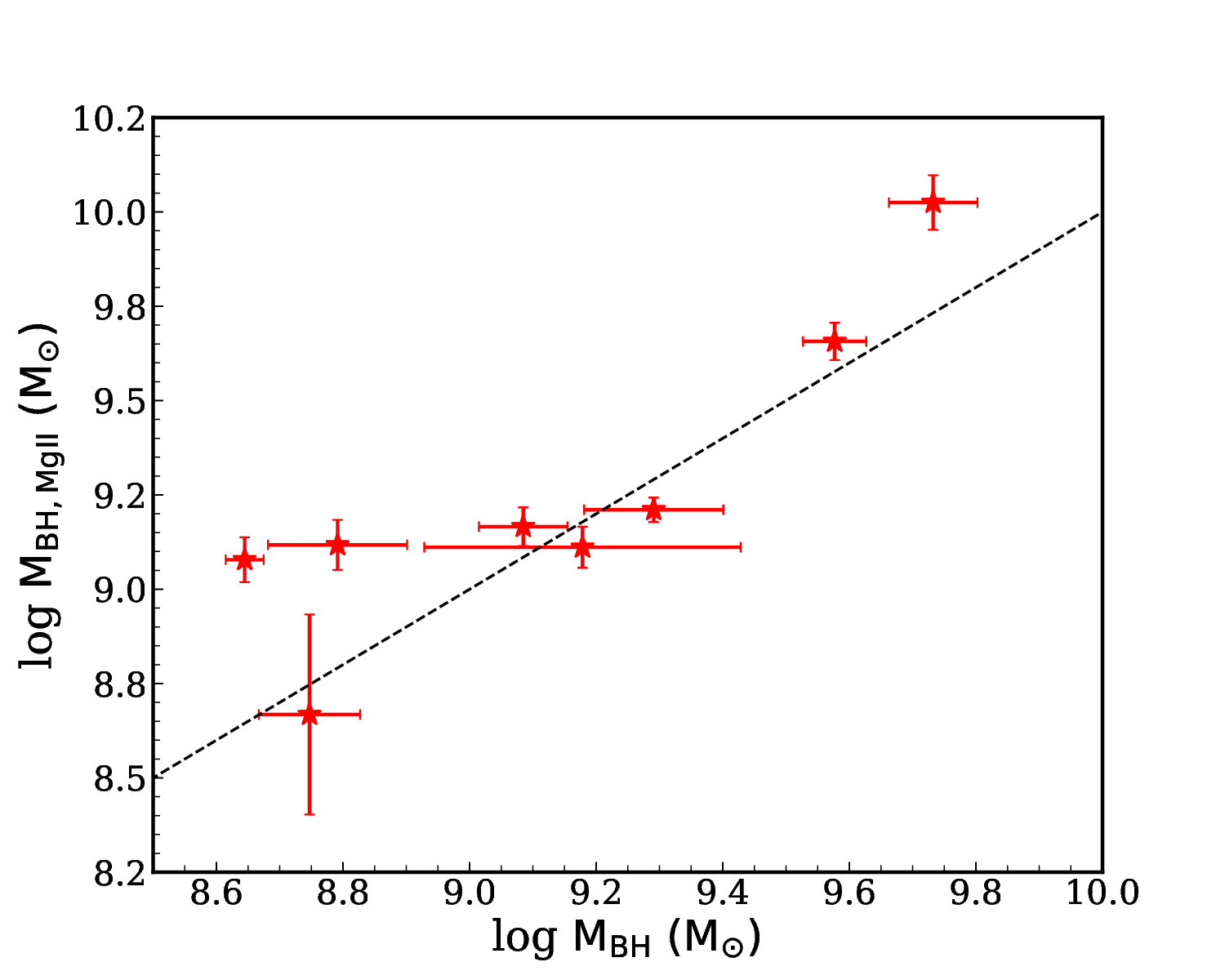}
\caption{Comparison of the H$\beta$-based virial SMBH masses with the \iona{Mg}{ii}-based virial SMBH masses for eight of the nine quasars; J0223 does not have a \iona{Mg}{ii}-based virial mass.
The error bars represent only measurement uncertainties.
The black line represents the line of equality.}
\label{fig-bhmc}
\end{figure}

%%% F6 l5100 vs. mbh/lamedd pic
\begin{figure}
\centering \includegraphics[width=\linewidth,keepaspectratio]{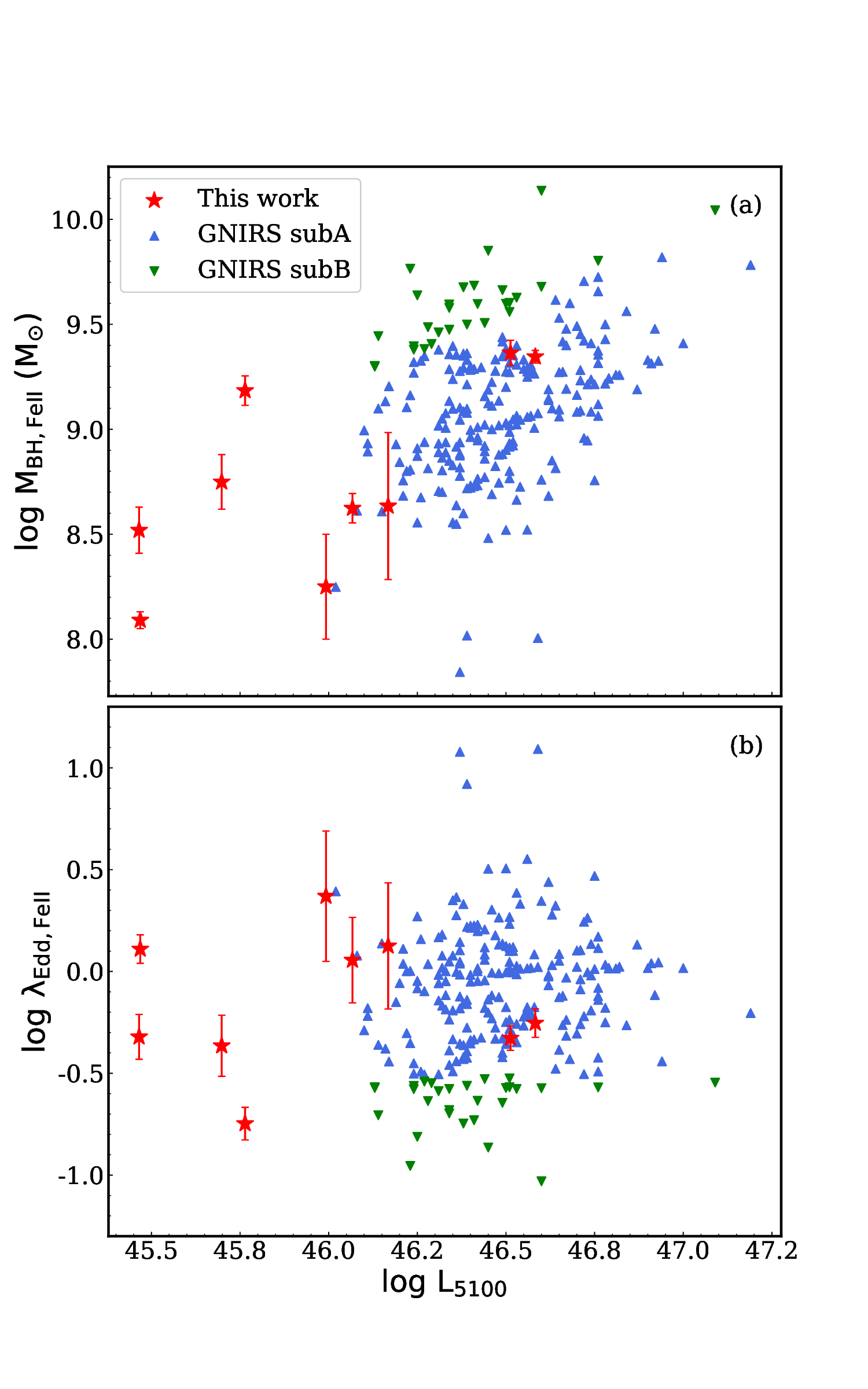}
\caption{(a) $M_{\rm BH,FeII}$ and (b) $\lambda_{\rm Edd,FeII}$ vs. monochromatic luminosity at 5100~\AA\ ($L_{\rm 5100}$).
Our sample is represented by the red stars, while the GNIRS subA and subB samples are shown as the blue and green triangles, respectively.
Our quasars show overall lower luminosities and smaller SMBH masses than the comparison samples.
But their $\lambda_{\rm Edd,FeII}$ distribution is similar to that of the full comparison sample.}
\label{fig-l5100}
\end{figure}

%%% F8 bhm vs. lamedd pic
\begin{figure}
\centering \includegraphics[width=\linewidth,keepaspectratio]{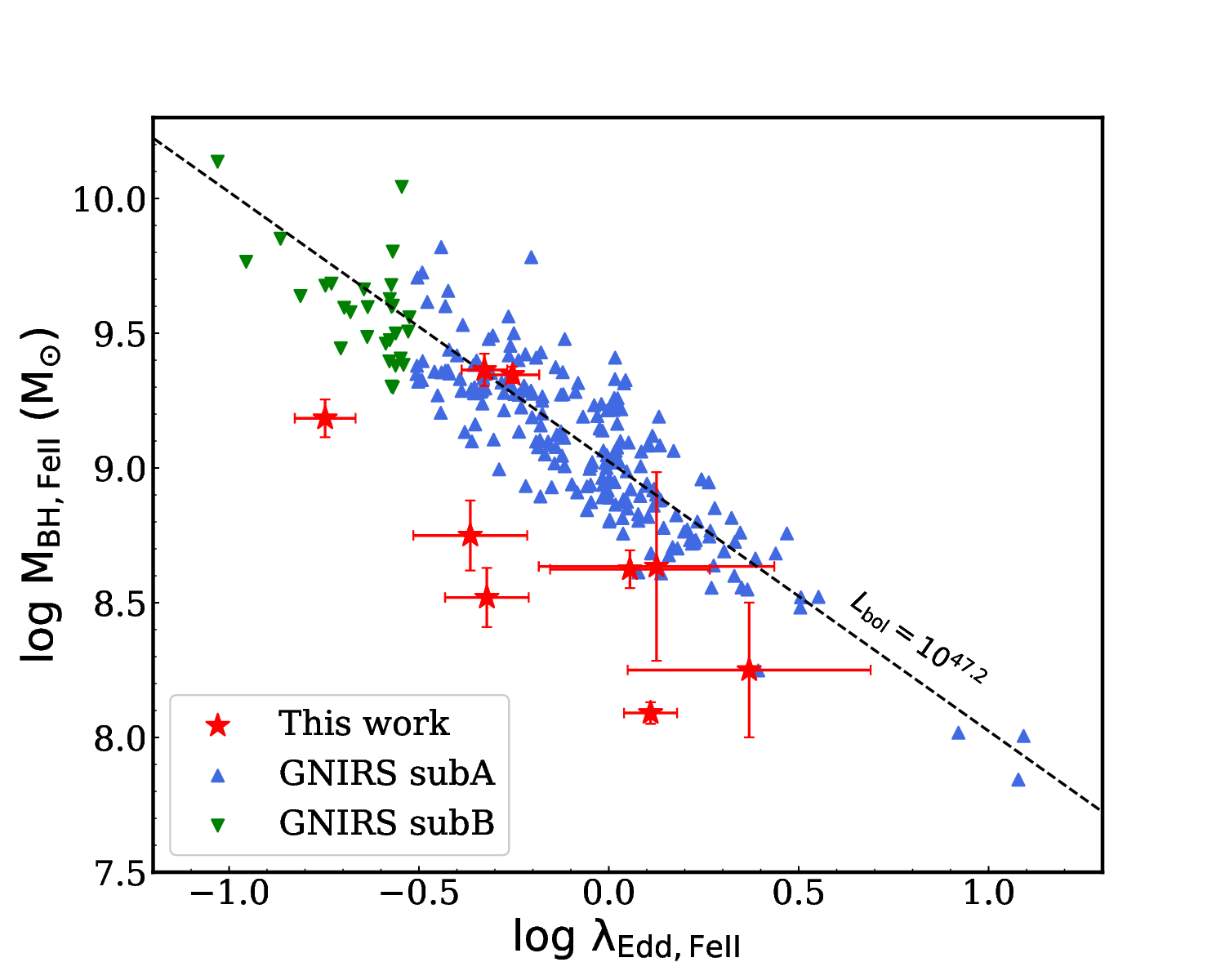}
\caption{$M_{\rm BH,FeII}$ vs. $\lambda_{\rm Edd,FeII}$ for our sample (red stars) and the comparison samples (blue and green triangles).
The dashed line corresponds to a constant bolometric luminosity of $10^{47.2}$ erg~s$^{-1}$, the median bolometric luminosity of the GNIRS sample.
There is no significant difference between the $\lambda_{\rm Edd,FeII}$ distributions of our sample and the GNIRS sample.}
\label{fig-bhmlam}
\end{figure}

%% black hole mass and ledd ratio table
\begin{deluxetable*}{lcccccc}
\tablewidth{0pt}
\tablecaption{Virial SMBH Masses and Eddington Ratios}
\tablehead{
\colhead{Source Name}&
\colhead{log~$M_{\rm BH}$}&
\colhead{log~$M_{\rm BH,MgII}$}&
\colhead{log~$M_{\rm BH,FeII}$}&
\colhead{log~$L_{\rm 5100}$}&
\colhead{$\rm \lambda_{Edd}$}&
\colhead{$\rm \lambda_{Edd,FeII}$} \\
\colhead{(SDSS J)}&
\colhead{($\rm M_{\odot}$)}&
\colhead{($\rm M_{\odot}$)}&
\colhead{($\rm M_{\odot}$)}&
\colhead{($\rm erg~s^{-1}$)}&
\colhead{}&
\colhead{} \\
\colhead{(1)}   &
\colhead{(2)}   &
\colhead{(3)}   &
\colhead{(4)}   &
\colhead{(5)}   &
\colhead{(6)}   &
\colhead{(7)}
}
\startdata
021830.59$-$045622.9 &$9.29\pm0.11$ & $9.21\pm0.03$  &$9.18\pm0.07$ & $45.76$ & $0.14\pm0.09$ & $0.18\pm0.08$ \\
022354.80$-$044815.0 &$8.95\pm0.16$ & --             &$8.25\pm0.25$ & $45.99$ & $0.47\pm0.11$ & $2.34\pm0.32$ \\
080612.03$+$194853.6 &$8.64\pm0.03$ & $9.08\pm0.06$  & $8.09\pm0.03$ & $45.47$ & $0.36\pm0.04$ & $1.29\pm0.05$ \\
083850.15$+$261105.4 &$9.58\pm0.05$ & $9.66\pm0.05$  &$9.35\pm0.03$ & $46.58$ & $0.33\pm0.04$ & $0.56\pm0.07$ \\
102117.74$+$131545.9 &$9.08\pm0.07$ & $9.17\pm0.05$  &$8.62\pm0.07$ & $46.07$ & $0.39\pm0.03$ & $1.14\pm0.21$ \\
104401.13$+$212803.9 &$9.18\pm0.25$ & $9.11\pm0.05$  &$8.63\pm0.35$ & $46.17$ & $0.38\pm0.16$ & $1.34\pm0.31$ \\
111518.58$+$531452.7 &$8.79\pm0.11$ & $9.12\pm0.07$  &$8.75\pm0.13$ & $45.70$ & $0.39\pm0.13$ & $0.43\pm0.15$ \\
121248.94$+$013727.1 &$8.75\pm0.08$ & $8.67\pm0.27$  &$8.52\pm0.11$ & $45.47$ & $0.28\pm0.08$ & $0.48\pm0.11$ \\
142435.97$+$421030.4 &$9.73\pm0.07$ & $10.02\pm0.07$ &$9.36\pm0.06$ & $46.51$ & $0.20\pm0.03$ & $0.47\pm0.06$ \\
\enddata
\tablecomments{Column (1): Name of the object.
Column (2): Logarithm of the SMBH mass derived from Equation~\ref{eq1}. The errors are measurement uncertainties.
Column (3): Logarithm of the SMBH mass based on the \iona{Mg}{ii} lines \citep{Shen2012}. 
Column (4): Logarithm of the SMBH mass derived from Equation~\ref{eq_dp}. 
Column (5): Logarithm of the continuum luminosity at rest-frame 5100~\AA. 
Column (6): Eddington ratio derived using $M_{\rm BH}$. 
Column (7): Eddington ratio derived using $M_{\rm BH,FeII}$.}
\label{tab:mbh}
\end{deluxetable*}

%%%%%%%%%%%%%%%%%%%%%%%%%%%%
\section{Results}
\label{sec:result}

%%%%%%%%%%%%%%%%%%%%%%%%%%%%
\subsection{SMBH Masses and Eddington Ratios}
\label{subsec:bhmledd}

With the P200 spectra, we are able to estimate H$\beta$-based single-epoch virial SMBH masses.
We first adopted the \citet{Vestergaard2006} formula that employs the traditional $R$--$L$ relation:
\begin{align}
	{\rm log}\left(\frac{M_{\rm BH}}{\rm M_{\odot}}\right) &= 0.91 + 0.5\left(\frac{L_{5100} }{10^{44}~{\rm erg~s^{-1}}}\right) \notag \\
    &\quad  + 2\left[\frac{{\rm FWHM} \left( {\rm H}\beta\right)}{{\rm km~s}^{-1}}\right].
\label{eq1}
\end{align}
The derived $\log(M_{\rm BH}/\rm{M_{\odot}})$ values are listed in Table~\ref{tab:mbh}, ranging from 8.64 to 9.73.
Measurement uncertainties are shown in Table~\ref{tab:mbh}, which are much smaller than the expected systematic uncertainties ($\rm \gtrsim 0.4$~dex; see Section~\ref{sec:intro}).
We then compared these masses to the \iona{Mg}{ii}-based virial masses ($M_{\rm BH,MgII}$).
We adopted Equation (3) of \citet{Shen2012} and the \iona{Mg}{ii} and 3000~\AA\ monochromatic luminosity ($L_{\rm 3000}$) measurements in the \citet{Wu2022} catalog to compute $M_{\rm BH,MgII}$,
also listed in Table~\ref{tab:mbh}.
J0223 does not have a $L_{\rm 3000}$ measurement in \citet{Wu2022}, and thus we do not include this quasar in the comparison.
The comparison of $M_{\rm BH}$ and $M_{\rm BH,MgII}$ is displayed in Figure~\ref{fig-bhmc}.
The two sets of SMBH masses agree well, with a median offset of $-0.08$ dex and a scatter of 0.19 dex.
Thus the \iona{Mg}{ii}-based SMBH masses are generally reliable, consistent with previous assessments \citep[e.g.,][]{Shen2012,Trakhtenbrot2012,Mejia2016,Wangshu2019}.

We then estimated H$\beta$-based single-epoch virial SMBH masses ($M_{\rm BH,FeII}$) with the updated $R$--$L$ relation in \cite{DU2019} that accounts for shorter reverberation mapping time lags observed in AGNs with high accretion rates \citep[e.g.,][]{Wang2014,Du2016,gravity2024}.
The formula is:  
\begin{align}
    \log\left(\frac{M_{\rm BH,FeII}}{\rm{M_{\odot}}}\right) &= 7.83 + 2~\log\left[\frac{{\rm FWHM}({\rm H}\beta)}{10^3~{\rm km~s}^{-1}}\right] \notag \\
    &\quad + 0.45~\log\left(\frac{L_{5100}}{10^{46}~{\rm erg~s^{-1}}}\right) - 0.35~R_{\rm Fe\ II,Flux}. %check again the define
\label{eq_dp}
\end{align}
In this formula, $R_{\rm Fe\ II,Flux}$ is defined similarly to $R_{\rm Fe\ II}$ in Footnote~2, but being a flux ratio instead of REW ratio, and the two quantities are generally very close to each other.
The resulting $M_{\rm BH,FeII}$ values are listed in Table~\ref{tab:mbh}, ranging from 8.09 to 9.36.
We caution that for objects with the H$\beta$ line best fitted with multiple broad components (e.g., J0838 and J1115), the FWHM(H$\beta$) value measured from the combined profile might be overestimated if the broad profile is affected by outflows/inflows and inaccuracies in continuum subtraction (including the \iona{Fe}{ii} pseudo continuum).
In this case, $M_{\rm BH,FeII}$ would be overestimated and the corresponding Eddington ratio would be underestimated.
By construction, the $M_{\rm BH,FeII}$ values are smaller than $M_{\rm BH}$, and for our sample, they are on average 0.36 dex smaller.
For quasars with smaller $R_{\rm Fe\ II}$ values (e.g., J0218), the differences between the two sets of masses are small.

\begin{figure}
\centering \includegraphics[width=\linewidth,keepaspectratio]{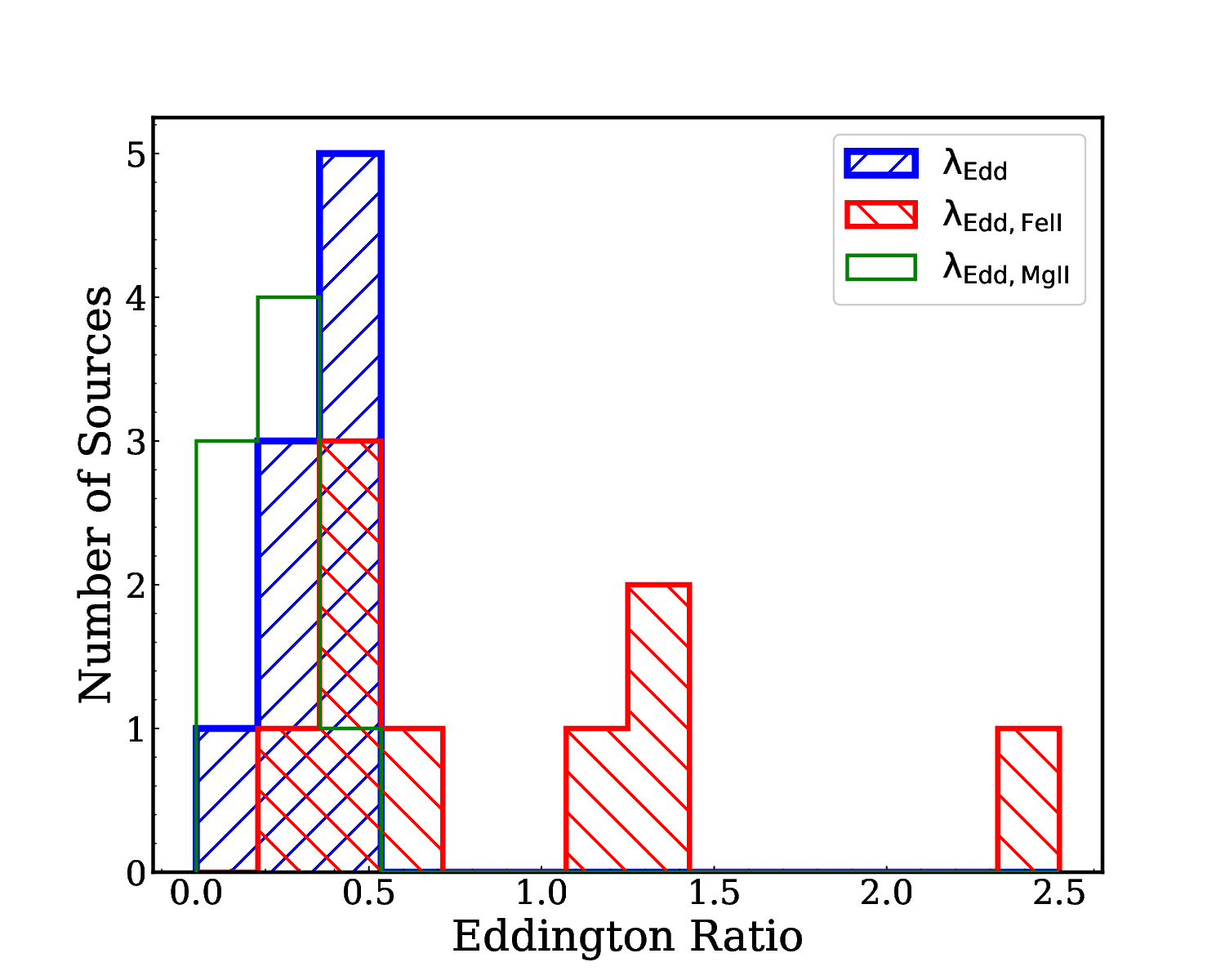}
\caption{Histograms showing the distributions of $\lambda_{\rm Edd}$ based on three different estimates of $M_{\rm BH}$: $M_{\rm BH}$, $M_{\rm BH,FeII}$, and $M_{\rm BH,MgII}$.}
\label{fig-mbhhist}
\end{figure}

To estimate bolometric luminosities, we adopted the luminosity-dependent correction factor in \citet{Netzer2019}: $k_{\rm bol}= 40\times[L_{\rm 5100}/10^{42}\rm \ erg\ s^{-1}]^{-0.2}$, where $L_{\rm 5100}$ is the 5100~\AA\ monochromatic luminosity derived in Section~\ref{subsec:nirfit} (listed in Table~\ref{tab:mbh}).
The two sets of Eddington ratios are then derived:
\begin{equation}
	\lambda_{\rm Edd} = \left[k_{\rm bol} \times L_{\rm 5100}\right] / \left[ 1.5\times 10^{38}~M_{\rm BH}/{\rm M_{\odot}}\right],
\label{eq2a}
\end{equation}
\begin{equation}
	\lambda_{\rm Edd,FeII} = \left[k_{\rm bol} \times L_{\rm 5100}\right] / \left[ 1.5\times 10^{38}~M_{\rm BH,FeII}/{\rm M_{\odot}}\right].
\label{eq2b}
\end{equation}
The results are listed in Table~\ref{tab:mbh}.
Our quasars have $\lambda_{\rm Edd}$ ($\lambda_{\rm Edd,FeII}$) ranging from 0.14 to 0.47 (0.18 to 2.34), with a median value of 0.36 (0.56), and six (eight) of the nine quasars have $\lambda_{\rm Edd}>0.3$ ($\lambda_{\rm Edd,FeII}>0.3$).
We show in Figure~\ref{fig-mbhhist} a comparison of different $\lambda_{\rm Edd}$ sets ($\lambda_{\rm Edd}$, $\lambda_{\rm Edd,MgII}$, $\lambda_{\rm Edd,FeII}$) derived from the three different estimates of $M_{\rm BH}$; $\lambda_{\rm Edd,FeII}$ values are the highest by design.

%%% gam vs. lambda-edd pic

We then derived $M_{\rm BH,FeII}$ and $\lambda_{\rm Edd,FeII}$ for the comparison samples.
The \citet{Matthews2023} catalog does not provide $R_{\rm Fe\ II,Flux}$ and thus we used $R_{\rm Fe\ II}$ in Equation~\ref{eq_dp}.
We show $M_{\rm BH,FeII}$ and $\lambda_{\rm Edd,FeII}$ versus $L_{5100}$ distributions for our sample and the comparison samples in Figure~\ref{fig-l5100}.
For statistical comparisons, we ran the Kolmogorov-Smirnov (K-S) test.
Our quasar sample exhibits systematically lower luminosities, with a median $L_{\rm bol}$ that is 0.37~dex lower than that of the comparison samples.
Additionally, the median $M_{\rm BH,FeII}$ is smaller by 0.51~dex compared to the comparison samples.
The K-S test results for the SMBH masses are listed in Table~\ref{tab:Samplestats}.
The K-S test results for $\lambda_{\rm Edd,FeII}$ (Table~\ref{tab:Samplestats}) indicate that our quasars have similar Eddington ratios as the GNIRS sample or the GNIRS subA sample; the difference between our sample and the GNIRS subB sample is simply a selection effect.
Therefore, the smaller SMBH masses of our objects are likely linked to the lower luminosities.
The $M_{\rm BH,FeII}$ versus $\lambda_{\rm Edd,FeII}$ distributions for our sample and the comparison samples are shown in Figure~\ref{fig-bhmlam}.
The GNIRS quasars have a narrow $L_{\rm Bol}$ range, and they are clustered along the median $L_{\rm Bol}$ ($10^{47.2}~\rm{erg~s^{-1}}$) line.
Our objects are disjoint from the GNIRS sample, toward a lower overall $L_{\rm Bol}$ value, but they distribute similarly in terms of $\lambda_{\rm Edd,FeII}$.  
As mentioned in Section~\ref{sec:intro}, the $\lambda_{\rm Edd}$ ($\lambda_{\rm Edd,FeII}$) parameter is not reliable for identifying super-Eddington accretion, and thus it is not surprising that we do not find significant difference between the $\lambda_{\rm Edd,FeII}$ distributions.

\begin{figure}
\centering \includegraphics[width=\linewidth,keepaspectratio]{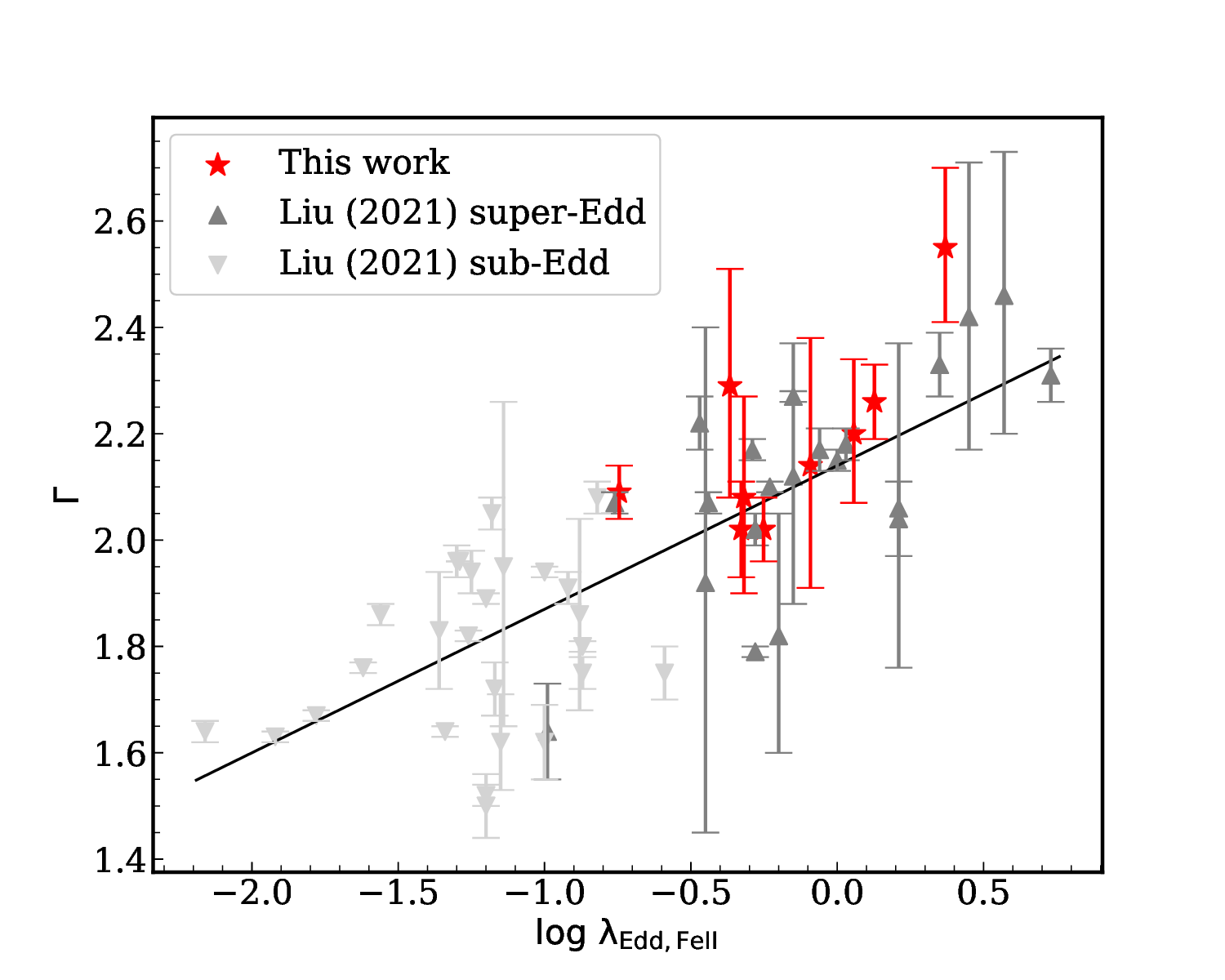}
\caption{$\Gamma$ vs. $\lambda_{\rm Edd,FeII}$ for our sample (red stars) and the super- (gray triangles) and sub-Eddington (lightgray triangles) accreting quasars in \citet{Liu2021}.
The best-fit relation from \citet{Liu2021} is displayed as the solid line.
Our quasars are generally consistent with the expectation from the \citet{Liu2021} relation.}
\label{fig-gamledd}
\end{figure}

Given the above $\lambda_{\rm Edd,FeII}$ measurements, we plot the distribution of our quasars in the $\Gamma$ versus $\lambda_{\rm Edd,FeII}$ plane in Figure~\ref{fig-gamledd}. 
The \citet{Liu2021} sample objects have reverberation-mapped $M_{\rm BH}$ values, and Equation~\ref{eq_dp} was calibrated based on such masses.
Therefore, there should be no systematic bias in the $M_{\rm BH}$ measurements.
Small offsets might arise from the $L_{\rm Bol}$ computations, as we adopted optical bolometric corrections while \citet{Liu2021} employed spectral energy distribution (SED) integration.
Overall, the two samples are comparable in terms of the parameter measurements, and our quasars do largely follow the best-fit relation from \citet{Liu2021}.
We also note that our $\Gamma$ selection may tend to pick up quasars scattered above the relation (a selection bias).

%%% gam vs. lambda-edd pic

%%% F5 other distributions pic
\begin{figure*}
\centering \includegraphics[width=\textwidth,keepaspectratio]{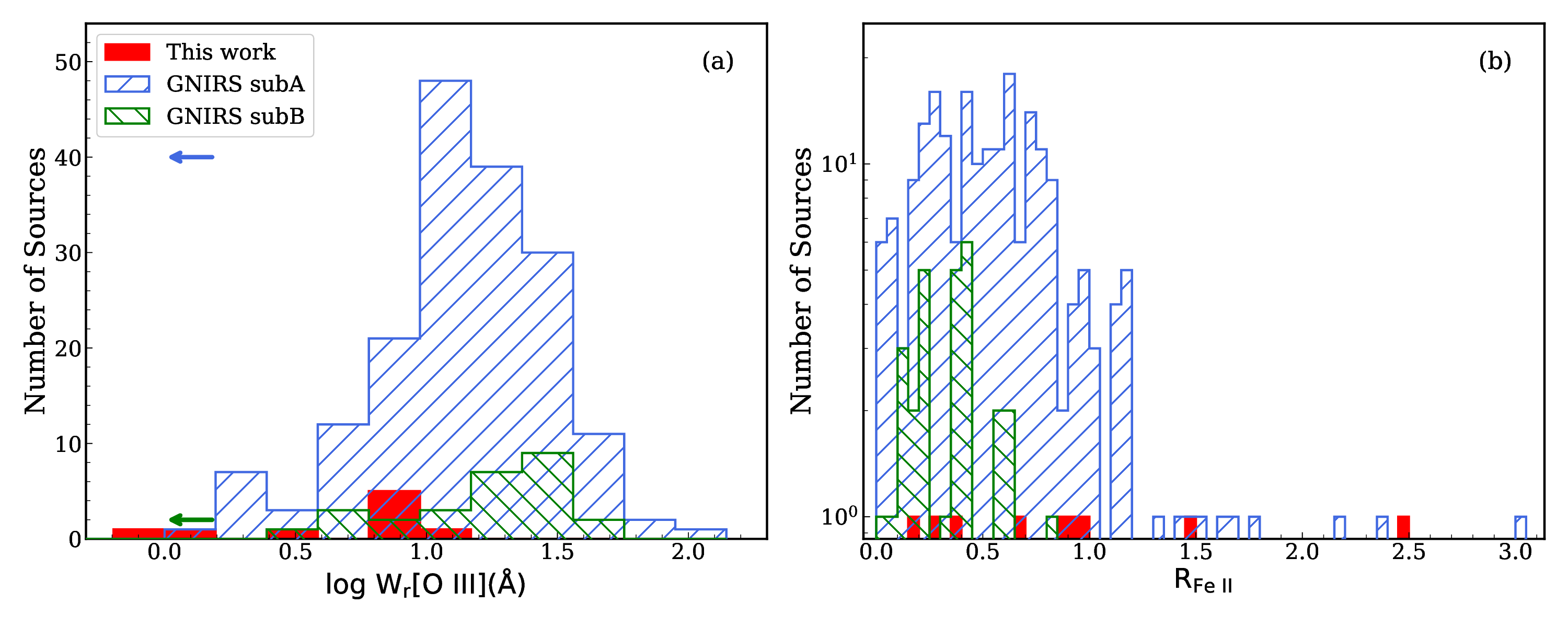}
\caption{Distributions of (a) log $W_{r}$[\iona{O}{iii}] and (b) $R_{\rm Fe\ II}$ for our sample (red solid histograms) and the GNIRS subA (blue histograms) and subB (green histograms) comparison samples.
For the $W_{r}$[\iona{O}{iii}] distributions, upper limits are depicted using leftward arrows with the y-axis value representing the numbers of sources.
Our objects show weaker [\iona{O}{iii}] and marginally stronger \iona{Fe}{ii} emission than the full GNIRS sample, and the differences are enhanced if the GNIRS subB sample is compared to.}
\label{fig-elp}
\end{figure*}

\begin{figure*}
\centering \includegraphics[width=\textwidth,keepaspectratio]{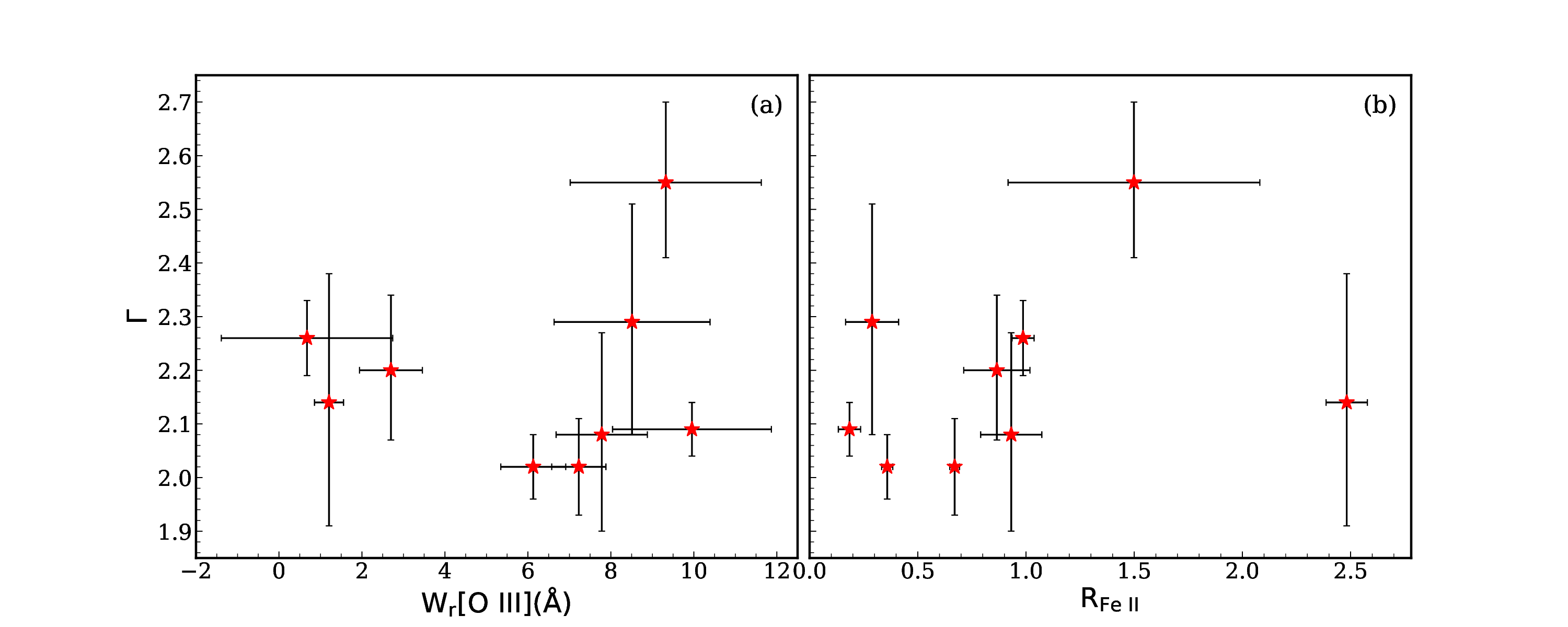}
\caption{$\Gamma$ vs. (a) $W_{r}$[\iona{O}{iii}] and (b) $R_{\rm Fe\ II}$ for our objects.
There is no clear negative correlation between $\Gamma$ vs. $W_{r}$[\iona{O}{iii}] or positive correlation between $\Gamma$ vs. $R_{\rm Fe\ II}$, probably affected by the limited sample size and the large $\Gamma$ errors.}
\label{fig-gam}
\end{figure*}

%%% comparison sample test table
\begin{deluxetable}{lcccccc}
\tablecaption{Statistical Test Results Between Our Sample and the GNIRS full, subA (super-Eddington), and subB (sub-Eddington) samples}
\tablehead{
\colhead{}&
\multicolumn{6}{c}{P-P test or K-S test} \\
\cmidrule(r){2-7}
\colhead{Comparison sample}&
\multicolumn{2}{c}{GNIRS}&
\multicolumn{2}{c}{GNIRS subA}&
\multicolumn{2}{c}{GNIRS subB}\\
\cmidrule(r){2-3} \cmidrule(r){4-5} \cmidrule(r){6-7}
\colhead{Parameter} & 
\colhead{Statistic} &  
\colhead{$P_{\rm null}$}&
\colhead{Statistic} &  
\colhead{$P_{\rm null}$}&
\colhead{Statistic} &  
\colhead{$P_{\rm null}$} \\
\colhead{(1)}   &
\colhead{(2)}   &
\colhead{(3)}   &
\colhead{(4)}   &
\colhead{(5)}   &
\colhead{(6)}   &
\colhead{(7)}
}
\startdata
$\rm W_{r}$[\iona{O}{iii}]& 2.67 &  0.0075 & 2.58 &0.0099 & 2.84 &0.0045 \\                      
$\rm R_{\rm Fe\ II}$    & 0.42 &  0.06 & 0.40 &  0.09 & 0.63 &  0.004 \\
$M_{\rm BH,FeII}$ &0.55 &0.006 & 0.53& 0.008 & 0.93 &6.7e-7 \\
$\lambda_{\rm Edd,FeII}$ &0.22 &0.71 & 0.31& 0.29 & 0.89 &3.0e-6 \\
\enddata
\tablecomments{Column (1): Name of the comparison parameter. Columns (2)--(7): P-P or K-S test results (statistic and null-hypothesis possibility).}
\label{tab:Samplestats}
\end{deluxetable}

%%%%%%%%%%%%%%%%%%%%%%%%%%%%
\subsection{Emission-Line Properties, Comparisons, and Multiwavelength Properties}
\label{subsec:xrayp}

Given the P200/TSpec measurements of the emission-line properties (Table~\ref{tab:elp}), we compare our objects to the comparison samples.
We focus below on the [\iona{O}{iii}]~$\lambda 5007$ REW ($W_{r}$[\iona{O}{iii}]) and $R_{\rm Fe\ II}$ parameters that are likely associated with $\lambda_{\rm Edd}$; we also checked the H$\beta$ REW and FWHM parameters, but our sample objects do not show atypical distributions.
The distributions of $W_{r}$[\iona{O}{iii}] and $R_{\rm Fe\ II}$ for our sample and the comparison samples are displayed in Figure~\ref{fig-elp}.
For statistical comparisons, we ran the Peto-Peto Generalized Wilcoxon (P-P) test in the Astronomy Survival Analysis package (ASURV; \citealt{Feigelson1985,Lavalley1992}) for the $W_{r}$[\iona{O}{iii}] distributions which contain censored data; for the $R_{\rm Fe\ II}$ distributions, we ran the K-S test.
Test results including the test statistics and null-hypothesis probabilities are listed in Table~\ref{tab:Samplestats}.

The $W_{r}$[\iona{O}{iii}] values for our objects range from 0.7 to 10.0~$\textup{\AA}$ (Table~\ref{tab:elp} and Figure~\ref{fig-elp}a), with a median value of 7.23~$\textup{\AA}$,
smaller than the average $W_{r}$[\iona{O}{iii}] of typical SDSS quasars, which is 13.2~$\textup{\AA}$ (\citealt{Vanden2001}).
Our X-ray selected super-Eddington candidates show noticeably weaker [\iona{O}{iii}] emission than the GNIRS sample ($P_{\rm null}=0.0075$).
They still show weaker [\iona{O}{iii}] emission than the super-Eddington comparison sample (GNIRS subA; $P_{\rm null}=0.0099$). The difference is enhanced when compared to the sub-Eddington comparison sample (GNIRS subB; $P_{\rm null}=0.0045$).

The $R_{\rm Fe\ II}$ values for our objects range from 0.2 to 2.5 (Table~\ref{tab:elp} and Figure~\ref{fig-elp}b), with a mean value of 0.92.
The $R_{\rm Fe\ II}$ distribution differs marginally from that of the full comparison sample ($P_{\rm null} = 0.06$), and the difference is smaller when compared to the GNIRS subA sample.
But our objects do show stronger \iona{Fe}{ii} emission compared to the GNIRS subB sample ($P_{\rm null} =0.004$).
We note the $\lambda_{\rm Edd,FeII}$ parameter was computed invoking $R_{\rm Fe\ II}$ (Section~\ref{subsec:bhmledd}), and thus the construction of the subA and subB samples separated at $\lambda_{\rm Edd,FeII}=0.3$ has bulit-in  $R_{\rm Fe\ II}$ dependence.
If we instead use $M_{\rm BH}$ and $\lambda_{\rm Edd}$ computed with the traditional $R$--$L$ relation to extract the subB sample (83 quasars), the  $R_{\rm Fe\ II}$ difference between our objects and the subB sample drops to $P_{\rm null} =0.05$.
We also tested constructing the subA sample with $\lambda_{\rm Edd,FeII}>1.0$, which in principle isolates quasars with higher accretion rates (more similar to our sample).
The differences between the resulting 92 subA quasars and our nine quasars do appear to decrease, with the $W_{r}$[\iona{O}{iii}] $P_{\rm null}$ value increasing from 0.0099 to 0.013 and the $R_{\rm Fe\ II}$ $P_{\rm null}$ value increasing from 0.09 to 0.31.
The significant $W_{r}$[\iona{O}{iii}] difference remains.
It is possible that, due to the limited sample size, our nine quasars happen to have higher accretion rates on average than the subA quasars with $\lambda_{\rm Edd,FeII}>1.0$.

We then explored if there is any dependence between the two EV1 parameters and $\Gamma$.
The $\Gamma$ versus $W_{r}$[\iona{O}{iii}] and $R_{\rm Fe\ II}$ distributions are displayed in Figure~\ref{fig-gam}.
In principle, if $\Gamma$ is indeed an accretion-rate indicator, we would expect sources with larger $\Gamma$ to show smaller $W_{r}$[\iona{O}{iii}] and larger $R_{\rm Fe\ II}$.
We performed the Kendall's $\tau$ test in ASURV to check for correlations, but there appear to be no significant dependence between $\Gamma$ versus $W_{r}$[\iona{O}{iii}] ($P_{\rm null}=0.83$) or $\Gamma$ versus $R_{\rm Fe\ II}$ ($P_{\rm null}=0.35$).
The lack of statistically significant correlations might be caused by our limited sample size and the large $\Gamma$ errors (the Kendall's $\tau$ test does not consider errors).

%%% ew_oiii vs. ew_civ pic
\begin{figure}
\centering \includegraphics[width=\linewidth,keepaspectratio]{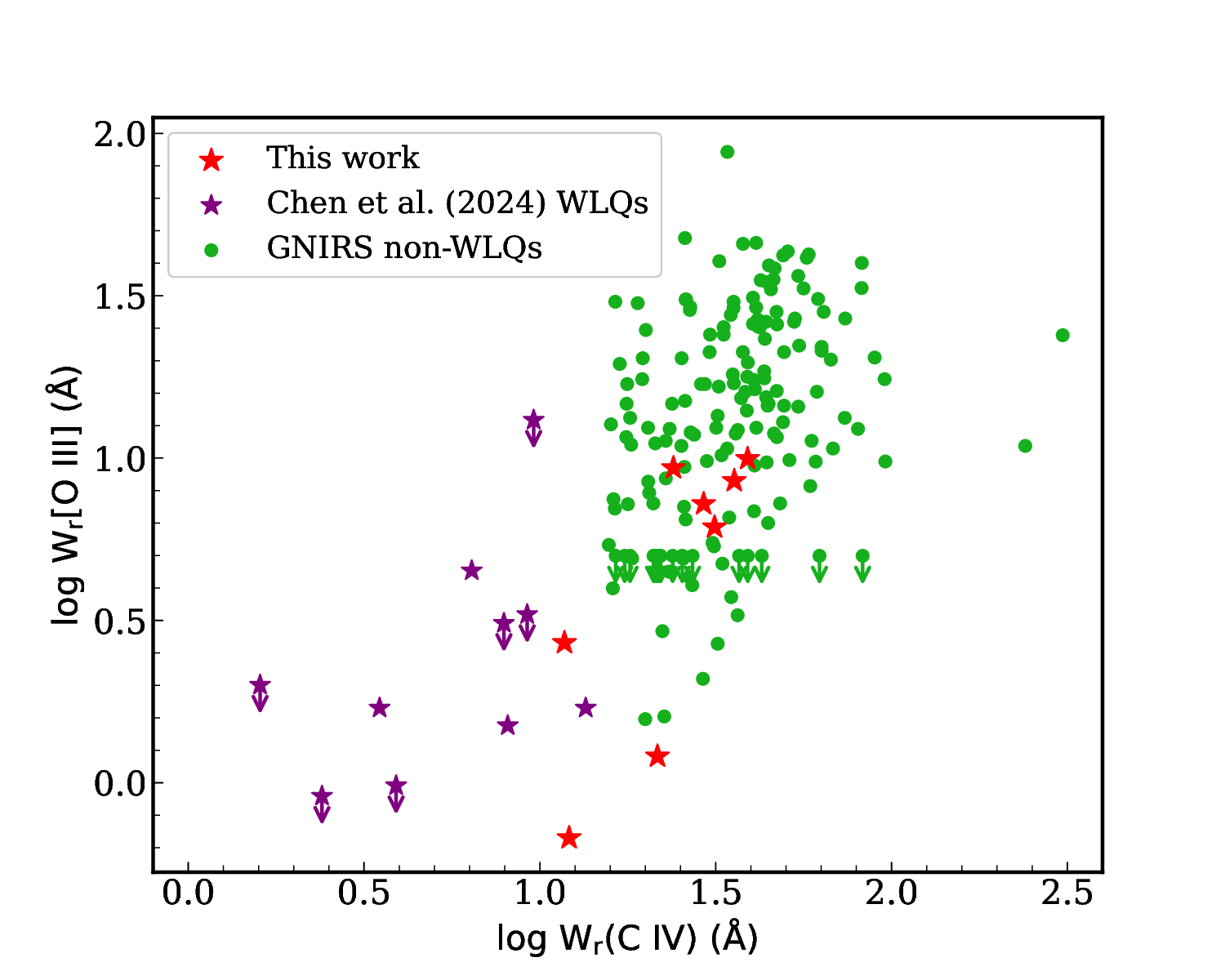}
\caption{log $W_{r}$[\iona{O}{iii}] vs. log $W_{r}$(\iona{C}{iv}) for our sample (red stars), the WLQs from \citet{Chen2024} (purple stars), and the GNIRS non-WLQs (green dots).
Upper limits are indicated by downward arrows.
WLQs predominantly occupy the bottom-left corner of the diagram.}
\label{fig-civoiii}
\end{figure}

We examined multiwavelength properties of our nine quasars.
We constructed optical-to-IR light curves using archival data from Zwicky Transient Facility Data Release 9 (ZTF DR9; \citealt{Bellm2019}) in the $zg$, $zr$, $zi$ bands, 
Panoramic Survey Telescope and Rapid Response System (Pan-STARRS) in the $g$, $r$, $i$, $z$, $y$ bands, 
and Near-Earth Object WISE (NEOWISE; \citealt{Mainzer2014}) in the W1 ($\rm 3.4~\mu m$), W2 ($\rm 4.6~\mu m$) bands.
We then examined the variability amplitudes of these light curves.
The maximum variability amplitudes are below $\approx0.5$ magnitude ($\approx60$\%), except for J1115 in the ZTF $zg$ band, which reaches $\approx0.8$ magnitude.
Some of the variability might be caused by changing seeing conditions in ground-based telescopes or flux calibration uncertainties.
Therefore, our quasars do not appear to show unusual optical-to-IR variability. 

We then compiled IR to X-ray SEDs for the nine quasars.
The details and the SEDs are presented in Appendix~\ref{sec:appendix2}.
Overall, our quasars show typical IR-to-UV SEDs.
They also display typical X-ray emission strength relative to their optical/UV emission (i.e., following the $\rm \alpha_{OX}$--$L_{2500~\textup{\AA}}$ relation).

%%%%%%%%%%%%%%%%%%%%%%%%%%%%
\section{Discussion}
\label{sec:dis}

%%%%%%%%%%%%%%%%%%%%%%%%%%%%
\subsection{X-ray Photon-Index Selection of Super-Eddington Accretion}
\label{subsec:gaccind}

As outlined in Section~\ref{sec:intro}, we can cross-verify the selected super-Eddington candidates using their Eddington ratio parameter, the optical [\iona{O}{iii}] and \iona{Fe}{ii} emission strengths, and other features relevant to super-Eddington accretion.
Our nine quasars have a $\lambda_{\rm Edd,FeII}$ distribution similar to the GNIRS comparison sample (Section~\ref{subsec:bhmledd}).
The lack of unusually high Eddington ratios does not contrast with super-Eddington accretion, as the Eddington ratio parameter is unreliable in the super-Eddington regime (Section~\ref{sec:intro}).
Low values of Eddington ratios (e.g., $<0.1$) might indicate sub-Eddington accretion, but our sample objects likely pass such thresholds.
A similar result was found in \citet{Chen2024}, where we compared properties of ten $z\sim 2$ weak emission-line quasars (WLQs) to the GNIRS comparison sample.
WLQs are selected to have exceptionally weak or no broad emission lines in the UV.
They also display higher occurrence rates of extreme X-ray weakness and X-ray variability.
The unusual multiwavelength properties of WLQs can be uniformly explained with super-Eddington accretion, and they might even be extremely super-Eddington (see discussion in Section~4.1 of \citealt{Chen2024}, and references therein).
However, the $\lambda_{\rm Edd}$ distribution of the \citet{Chen2024} WLQs, with a range of 0.18 to 1.73, is still similar to that of the GNIRS sample. 
Therefore, it is not surprising that we do not find significant difference between the Eddington ratio distributions of our sample and the GNIRS sample.

On the other hand, our quasars do show weaker [\iona{O}{iii}] emission and marginally stronger \iona{Fe}{ii} emission than the GNIRS sample.
The difference is enhanced when compared to the sub-Eddington subB sample.
We caution that the GNIRS sample is unlikely a representative sample of $z\sim 2$ quasars, as these flux-limited objects are biased toward high luminosities and probably high accretion rates.
Nevertheless, these EV1 trends are consistent with the expectation that the selected $\Gamma>2$ quasars have higher accretion rates than the GNIRS quasars.
We do not observe dependences between $\Gamma$ and the [\iona{O}{iii}] or \iona{Fe}{ii} strengths (Section~\ref{subsec:xrayp} and Figure~\ref{fig-gam}), which might be affected our limited sample size and the large $\Gamma$ errors.

The \citet{Chen2024} WLQs also show weaker [\iona{O}{iii}] emission and stronger \iona{Fe}{ii} emission than the GNIRS quasars,
and the differences are more significant given the smaller test null-hypothesis probabilities.
We list in Table~\ref{tab:elp} the \iona{C}{iv} REWs (\citealt{Wu2022}) of our quasars.
Two of them can be classified as WLQs with \iona{C}{iv} REWs $<15~\textup{\AA}$ (J1021 and J1044), and they show clearly weak [\iona{O}{iii}] emission with REWs of 2.7 and 0.7~\AA.
These results are consistent with WLQs being at the extreme end of the super-Eddington population.
We show in Figure~\ref{fig-civoiii} the locations of our quasars (eight objects with \iona{C}{iv} coverage) in the log $W_{r}$[\iona{O}{iii}] versus log $W_{r}$(\iona{C}{iv}) plane.
For comparison, the WLQs from \citet{Chen2024} and the GNIRS non-WLQs are also displayed.
$W_{r}$[\iona{O}{iii}] and $W_{r}$(\iona{C}{iv}) appear to be positively correlated, especially at the low $W_{r}$(\iona{C}{iv}) end.
Our quasars generally follow this trend, and they appear to bridge the WLQs and GNIRS non-WLQs.
The weak [\iona{O}{iii}] and \iona{C}{iv} emission in WLQs can be uniformly explained by the wind/disk obscuration scenario (see Section~4.1 of \citealt{Chen2024}).

Overall, the EV1 results from the P200 spectra appear promising, and X-ray photon-index selection of large samples of super-Eddington accreting quasars from the Chandra and XMM-Newton archives might indeed be feasible.
However, as introduced in Section~\ref{sec:intro}, the $\Gamma$--$\lambda_{\rm Edd}$ correlation itself is not necessarily robust.
A larger sample than our current one is needed to further assess the reliability of the X-ray selection of super-Eddington accreting quasars, which in turn might provide indirect support for the positive $\Gamma$--$\lambda_{\rm Edd}$ correlation.

%%% loiii vs. lx pic
\begin{figure}
\centering \includegraphics[width=\linewidth,keepaspectratio]{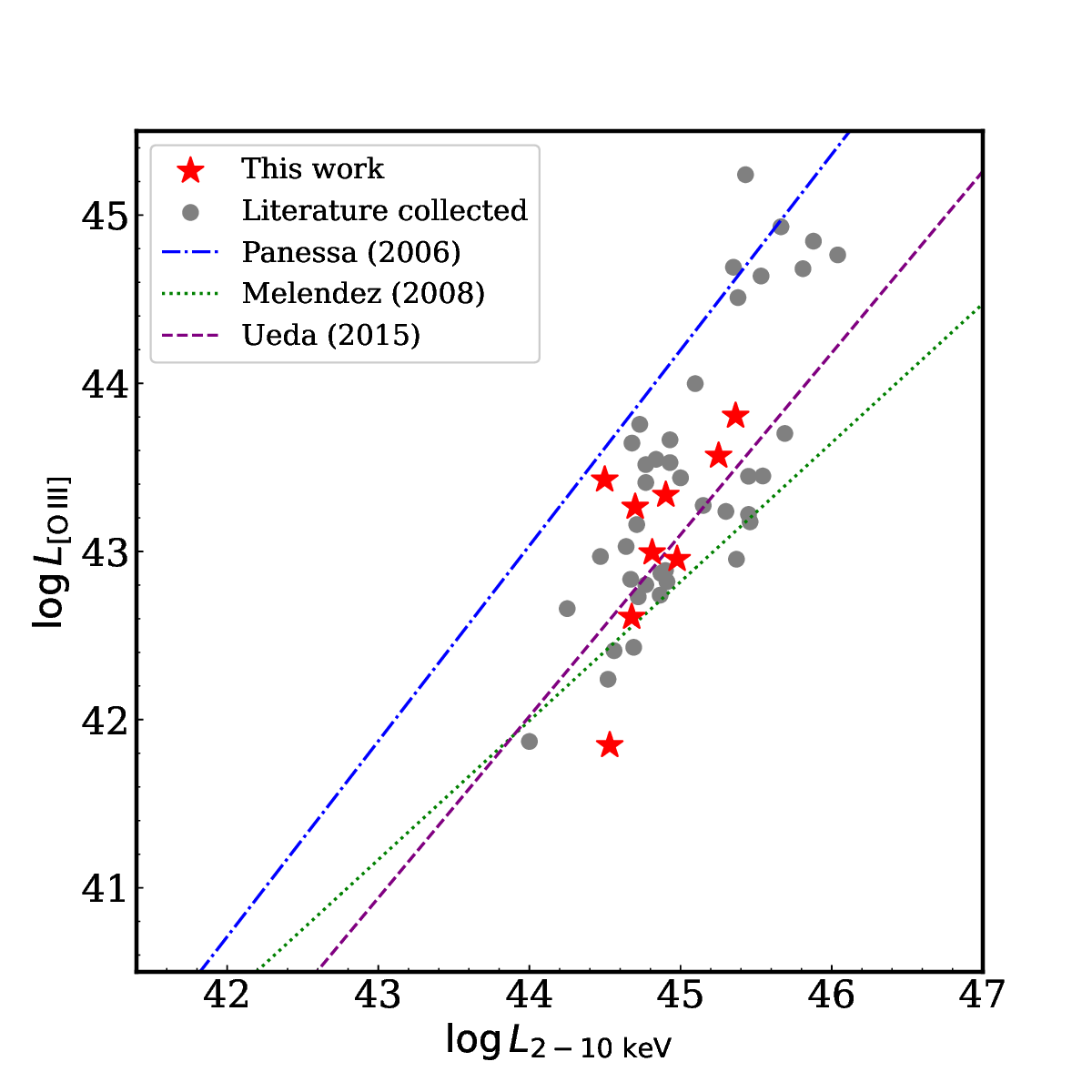}
\caption{$\log\,L_{\rm [O\,III]}$ vs. $\log\,L_{\rm 2-10~keV}$ for our targets (red stars) and data collected from literature (gray dots). 
The dot-dashed, dotted, and dashed lines represent the best-fit relations from 
\citet{Panessa2006}, \citet{Melendez2008}, and \citet{Ueda2015}, respectively.
Our quasars do not display significantly weaker $L_{\rm [O\,III]}$ compared to the other quasars with similar X-ray luminosities.}
\label{fig-lxloiii}
\end{figure}

As mentioned in Section~\ref{sec:intro}, super-Eddington accreting quasars also show much higher occurrence rates of extreme X-ray weakness and X-ray variability.
Our quasars have nominal X-ray emission strengths given their SEDs in Figure~\ref{fig-sed}.
We selected bright X-ray sources in the sample construction (Section~\ref{subsec:sample}), and thus they tend to be X-ray normal.
We do not observe significant X-ray variability either.
The estimated fraction of super-Eddington accretion quasars exhibiting extreme X-ray variability is $\sim 20\%$ (e.g., \citealt{Liu2019}).
We have a limited number of sample objects ($\approx4$) with well separated multi-epoch X-ray observations.
Thus it is within expectation that we do not observe significant X-ray variability.

We also examined our quasars in the $L_{\rm [O\,III]}$ versus $L_{\rm 2-10~keV}$ plane (Figure~\ref{fig-lxloiii}).
For comparison, high-redshift literature data \citep[e.g.,][]{Vietri2018,Zappacosta2020,Kakkad2020,Laurenti2022} are also shown.
Our quasars do not display significantly weaker $L_{\rm [O\,III]}$ compared to the other quasars with similar X-ray luminosities.
However, we note that some of these literature quasars are also highly accreting \citep[e.g.,][]{Vietri2018,Zappacosta2020,Laurenti2022} and some exhibit low $W_{r}$[\iona{O}{iii}] values \citep[e.g.,][]{Trefoloni2023}.
We also show in Figure~\ref{fig-lxloiii} three $L_{\rm [O\,III]}$--$L_{\rm 2-10~keV}$ relation lines from the literature, which were determined from AGN samples with lower luminosities.
Given the large scatter of $L_{\rm [O\,III]}$ at a given $L_{\rm 2-10~keV}$ for the literature data and the large dispersion of the three relations at the high-luminosity end, it is difficult to determine whether our quasars have lower-than-expected $L_{\rm [O\,III]}$.
It is worth noting that $W_{r}$[\iona{O}{iii}] is negatively correlated with quasar luminosity \citep[e.g.,][]{Sulentic2004,Netzer2006,Stern2012,Shen2016,Coatman2019}, opposite to the trend for $L_{\rm [O\,III]}$.
Therefore, $L_{\rm [O\,III]}$ might not be a good indicator of [\iona{O}{iii}] strength, as it is largely controlled by the distance/luminosity of the quasars; $L_{\rm [O\,III]}$ does span a much broader range than $W_{r}$[\iona{O}{iii}] (cf. Figure~\ref{fig-elp}).

%%%%%%%%%%%%%%%%%%%%%%%%%%%%
\subsection{An \texorpdfstring{[\iona{O}{iii}]}{oiii} Outflow Candidate}
\label{subsec:outflow}

Super-Eddington accreting quasars are expected to produce powerful accretion-disk winds \citep[e.g.,][]{Giustini2019,Jiang2019}, which may develop into massive galactic-scale outflows.
One source, J0838, displays significant outflow features characterized by blueshifted broad [\iona{O}{iii}] doublet emission lines (Figure~\ref{fig-spec1}).
The FWHMs and velocity offsets of the [\iona{O}{iii}]~$\lambda 4959$ and [\iona{O}{iii}]~$\lambda 5007$ lines are tied, and the line flux ratio is also fixed at the fiducial 1:3 ratio.
The resulting FWHM is $\rm = 1960\pm190~km~s^{-1}$ and the blueshift is $\rm = 690\pm170~km~s^{-1}$.
Previous reports of blueshifted broad [\iona{O}{iii}] emission lines in quasars are limited \citep[e.g.,][]{Brusa2015,Brusa2016,Zakamska2016,Bischetti2017,Xu2020,Fukuchi2023}.
These lines suggest strong [\iona{O}{iii}] outflows in the host galaxy, providing compelling evidence for quasar feedback.
Considering the high SNR (approximately 56 in the H$\beta$ region) of the P200/TSpec spectrum for J0838,
the possibility of strong \iona{Fe}{ii} contamination to the broad [\iona{O}{iii}] lines \citep[e.g.,][]{Kova2010,Bischetti2017} is low.
Therefore, we consider J0838 a good candidate with a strong [\iona{O}{iii}] outflow.

%%%%%%%%%%%%%%%%%%%%%%%%%%%%
\section{Summary and Future Work}
\label{sec:sumf}

Given the importance of super-Eddington accretion and the challenges in identifying such AGNs,
we tested the X-ray photon-index selection using the Chandra and XMM-Newton archival data.
We obtained P200/TSpec NIR spectra for a pilot sample of nine $\Gamma=2.0$--2.6 quasars at $z\approx1.4$--2.5, and we constructed comparison samples from the flux-limited GNIRS-DQS quasar sample at $z\approx1.5$--3.5 with NIR spectroscopy.
The key findings are as follows:

\begin{enumerate}
\item
Based on the P200 spectra, we derived H$\beta$-based single-epoch virial SMBH masses with the updated $R$--$L$ relation in \citet{DU2019} that accounts for shorter reverberation mapping time lags observed in AGNs with high accretion rates.
The resulting SMBH masses range from 8.09 to 9.36.
The corresponding Eddington ratios range from 0.18 to 2.34.
Our quasars show overall lower luminosities and smaller SMBH masses than the comparison samples.
But their Eddington ratio distribution is overall consistent with that of the GNIRS comparison sample.
The Eddington ratio parameter is not reliable for identifying super-Eddington accretion, which might explain the lack of difference.
See Section~\ref{subsec:bhmledd}.
\item
Our quasars show noticeably weaker [\iona{O}{iii}] emission than the GNIRS sample ($P_{\rm null}=0.0075$).
They show marginally stronger \iona{Fe}{ii} emission than the GNIRS sample ($P_{\rm null}=0.06$).
These results involving the EV1 parameters are consistent with the expectation that the selected $\Gamma>2$ quasars have higher accretion rates than the GNIRS quasars.
Our quasars show typical IR-to-X-ray SEDs, and they do not display unusual variability.
See Section~\ref{subsec:xrayp}.
\item
We report one object with clear broad and blueshifted [\iona{O}{iii}]~$\lambda 4959$ and [\iona{O}{iii}]~$\lambda 5007$ lines.
The line FWHM is $\rm = 1960\pm190~km~s^{-1}$ and the blueshift is $= 690\pm170$~km~s$^{-1}$.
The lines probably originate from a strong [\iona{O}{iii}] outflow that is driven by super-Eddington accretion.
See Section~\ref{subsec:outflow}.
\end{enumerate}

Overall, the X-ray photon-index selection of super-Eddington accreting quasars appears promising.
But a larger sample than our current one is needed to assess further the reliability of the selection.
We are planning to obtain P200 spectra for a larger sample ($\approx20$) of similar quasars, and then examine their Eddington ratios, [\iona{O}{iii}] and \iona{Fe}{ii} emission strengths (as well as the $\Gamma$ dependence of these EV1 parameters), X-ray emission strengths and variability.
Upon verification of the technique, it can also be applied to the eROSITA $+$ SDSS catalogs.
The eROSITA/eRASS1 catalog \citep{Merloni2024} provides X-ray measurements for $\approx2400$ $z>2$ SDSS DR16 quasars.

Another interesting direction to explore is to select super-Eddington accreting candidates with the EV1 parameters, i.e., 
small [\iona{O}{iii}] REWs or large $R_{\rm Fe\ II}$ values.
This technique may in principle be applied to high-redshift AGNs, such as those detected by JWST; the contribution from super-Eddington accretion is expected to be high in the early SMBH growth.
Since the physical nature for these EV1 treads remains unclear, cross-verifications from, e.g., X-ray photon indices, are needed to help understand any contribution from contaminants.

~\\

We thank the referee for providing helpful comments.
Y.C. and B.L. acknowledge financial support from the National Natural Science Foundation of China grant No. 12573016.
This research uses data obtained through the Telescope Access Program (TAP), which has been funded by the Strategic Priority Research Program “The Emergence of Cosmological Structures” (Grant No. XBD09000000), National Astronomical Observatories, Chinese Academy of Sciences, and the Special Fund for Astronomy from the Ministry of Finance. 
Observations obtained with the Hale Telescope at Palomar Observatory were obtained as part of an agreement between
the National Astronomical Observatories, Chinese Academy of
Sciences, and the California Institute of Technology.
This paper employs a list of Chandra datasets, obtained by the Chandra X-ray Observatory, contained in~\dataset[DOI: 10.25574/cdc.404]{https://doi.org/10.25574/cdc.404}.

%%%%%%%%%%%%%%%%%%%%%%%%%%%%
% reference
\bibliographystyle{aasjournal}
\bibliography{ms}

%%%%%%%%%%%%%%%%%%%%%%%%%%%%
\clearpage
\appendix

\section{X-ray properties of the other 12 $\Gamma>2$ quasars}
\label{sec:appendix1}

\begin{figure}
    \figurenum{A1}
    \centering \includegraphics[width=\linewidth]{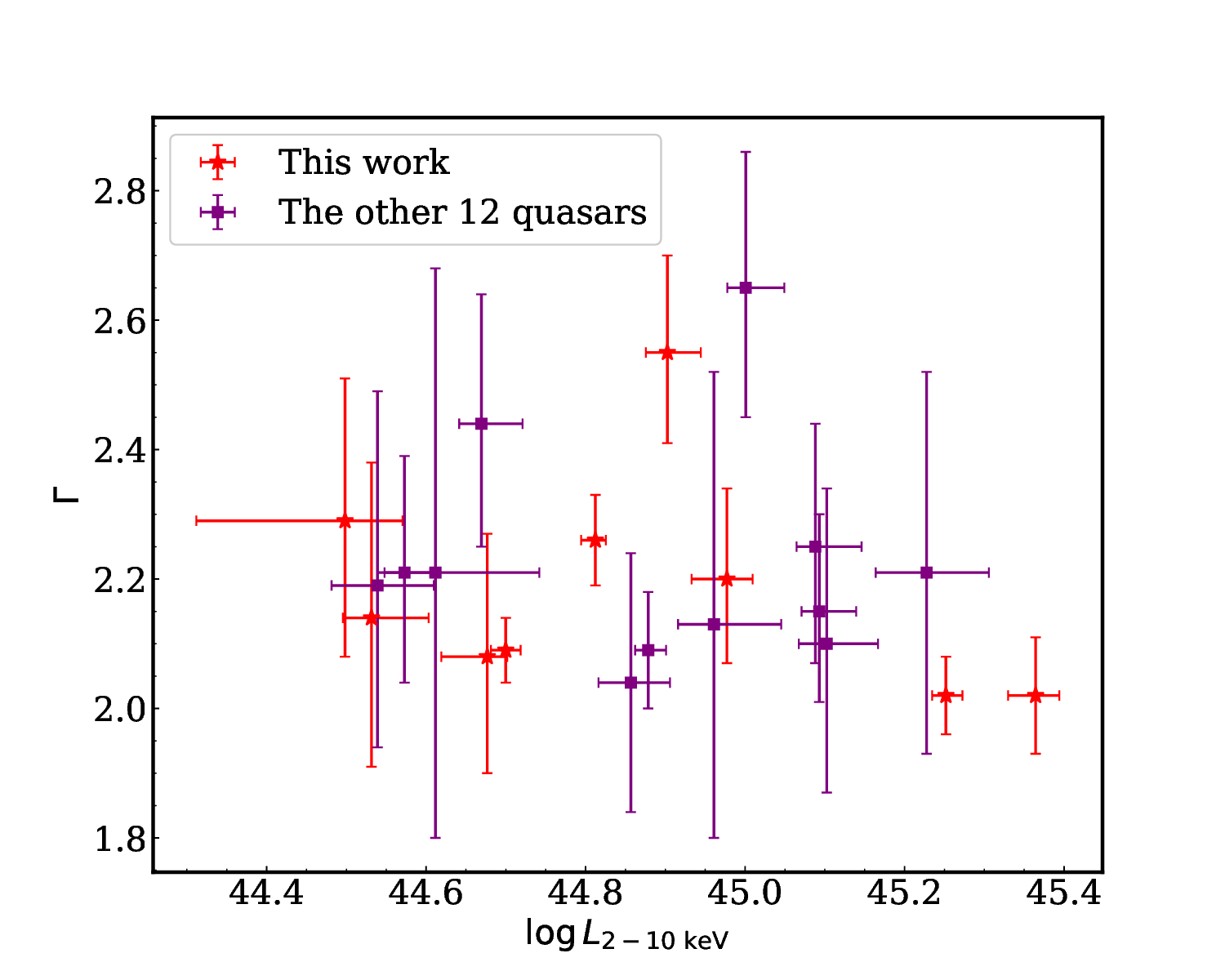}
    \caption{The $\Gamma$ vs. $L_{\rm 2-10~keV}$ distributions for the 21 $\Gamma>2$ quasars. 
    Our final sample objects (nine quasars) are in red, and the other 12 quasars without NIR observations are in purple.
    The $\Gamma$ and $L_{\rm 2-10~keV}$ distributions are overall consistent between the two groups.}
    \label{fig-Lgam}
\end{figure}

We present the X-ray properties of the other 12 ($21-9$) $\Gamma>2$ quasars without NIR observations in Table~\ref{tab:x_ray_12}.
The $\Gamma$ versus $L_{\rm 2-10~keV}$ distributions for all the 21 quasars are displayed in Figure~\ref{fig-Lgam}.
Overall, the nine targets in our sample do not exhibit unusual X-ray properties compared to the other 12 quasars.

\section{SEDs for our nine quasars}
\label{sec:appendix2}

\begin{figure*}
    \figurenum{B1}
    \centering \includegraphics[scale=0.33,keepaspectratio]{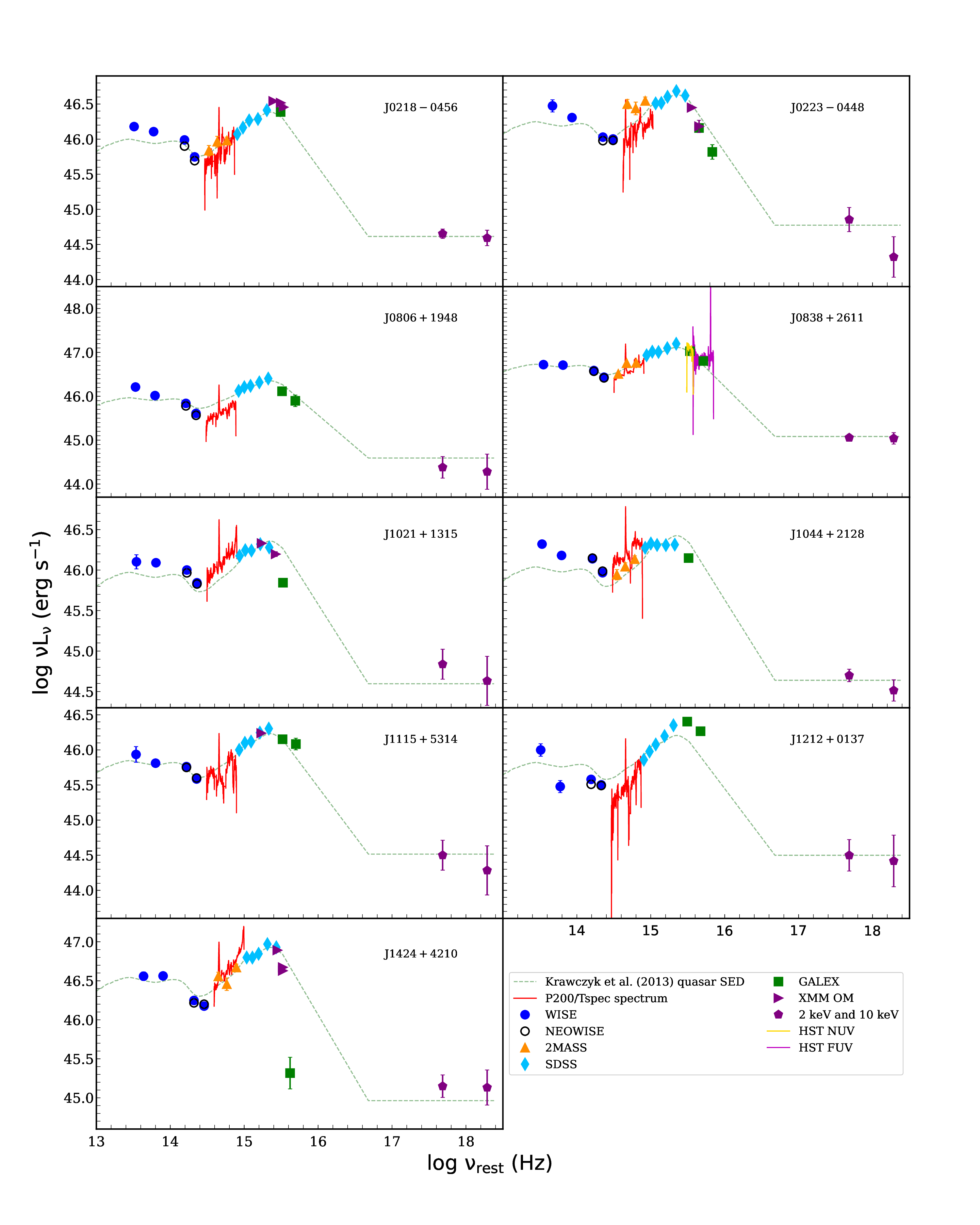}
    \caption{IR-to-X-ray SEDs for our sources.
    The IR--UV photometric data were collected from {WISE}, {NEOWISE}, 2MASS, SDSS, {GALEX}, and XMM-Newton OM.
    For NEOWISE, average measurements are displayed.
    Our P200/Tspec spectra are also included as red curves; for J0838, the HST UV spectra are displayed.
    Some of the error bars are smaller than the symbol size.
    The green dashed curve illustrates the mean quasar SED from \citet{Krawczyk2013}, normalized at the 2500~\AA\ luminosity interpolated from the SDSS photometric data.
    The X-ray component of the mean quasar SED is determined from the $\alpha_{\rm OX}$--$L_{\rm 2500~{\textup{\AA}}}$ relation in \cite{Steffen2006}.
    These photometric and spectroscopic data might be affected by mild variability effects.
    Overall, our quasars show typical IR-to-X-ray SEDs.}
    \label{fig-sed}
\end{figure*}

We show the IR to X-ray SEDs for the nine quasars in Figure~\ref{fig-sed}.
The IR to UV photometric data were gathered from the WISE, NEOWISE, 2MASS, {Galaxy Evolution Explorer} ({GALEX}; \citealt{Martin2005}), and SDSS (\citealt{York2000}) catalogs.
The X-ray luminosities at 2~keV and 10~keV derived from the best-fit models (Section~\ref{subsec:xrayfit}) along with available XMM-Newton OM photometric measurements were also included.
Additionally, we displayed the P200/Tspec spectra in the SEDs.
For J0838, we also added near-UV (NUV) and far-UV (FUV) spectra from the Hubble Space Telescope/Cosmic Origins Spectrograph (HST/COS).
The optical and UV data were corrected for Galactic extinction as described in Section~\ref{subsec:nirfit}.
For comparison, we plotted the mean SED of high-luminosity radio-quiet quasars in \citet{Krawczyk2013} normalized at the 2500~\AA\ luminosity.
The X-ray component of the mean quasar SED is determined from the $\rm \alpha_{OX}$--$L_{2500~\textup{\AA}}$ relation in \citet{Steffen2006}.
Mild variability is present between these multi-epoch photometric and spectroscopic data.
Overall, our quasars show typical IR-to-X-ray SEDs.

\tablewidth{0pt}
\begin{deluxetable*}{lccccccccccc}
\tablenum{A1}
\tablecaption{X-ray Observations Log}
\tablehead{
\colhead{Source Name}&
\colhead{$z_{\rm dr16}$}&
\colhead{Observatory}&
\colhead{Obs. Date}&
\colhead{Observation}&
\colhead{Exp. Time}&
\colhead{Net Counts}&
\colhead{$N_{\rm H, Gal}$}&
\colhead{$\Gamma$}&
\colhead{$L_{\rm 2-10~keV}$}
\\  
\colhead{(SDSS J)}& 
\colhead{} &
\colhead{} &
\colhead{} &
\colhead{ID}&
\colhead{(ks)} &
\colhead{} &
\colhead{$\rm (10^{20}~{\rm cm}^{-2})$} &
\colhead{} & 
\colhead{($\rm 10^{44}~{\rm erg}~~{\rm s}^{-1}$)} 
\\
\colhead{(1)}   &
\colhead{(2)}   &
\colhead{(3)}   &
\colhead{(4)}   &
\colhead{(5)}   &
\colhead{(6)} &
\colhead{(7)}   &
\colhead{(8)} &
\colhead{(9)} &
\colhead{(10)}
} 
\startdata
080433.98+303030.0 & 2.244 & XMM-Newton &2018-04-29& 0802220201 & 36.7 & 396/186/113 & 3.70 & $2.13_{-0.33}^{+0.39}$ & $9.14_{-0.90}^{+1.96}$  \\
084943.71+450024.2 & 1.599 & Chandra & 2000-05-04&927   & 119.2 & 727  & 2.70 & $2.09_{-0.09}^{+0.09}$ & $7.56_{-0.28}^{+0.40}$  \\
095908.68+025423.9 & 1.567 & Chandra & 2013-01-01&15213 & 48.9 & 176 & 1.84 & $2.44_{-0.19}^{+0.20}$ & $4.67_{-0.29}^{+0.59}$  \\
112104.93+432141.6 & 2.030   & Chandra & 2005-01-11&5771  & 19.8 & 121 & 2.75 & $2.10_{-0.23}^{+0.24}$ & $12.66_{-0.98}^{+2.02}$  \\
115911.43+440818.3 & 1.444 & Chandra & 2004-02-03&4739  & 71.6 & 278 & 1.32 & $2.21_{-0.17}^{+0.18}$ & $3.74_{-0.28}^{+0.29}$  \\
120249.34+554148.1 & 1.417 & Chandra & 2012-08-09&14026 & 48.0 & 452 & 1.13 & $2.21_{-0.41}^{+0.47}$ & $4.09_{-0.56}^{+1.43}$  \\
123530.35+262058.0 & 1.484 & Chandra & 2003-02-04&2884  & 28.0 & 190 & 1.30 & $2.25_{-0.18}^{+0.19}$ & $12.25_{-0.65}^{+1.75}$  \\
124220.07+023257.6 & 2.220   & XMM-Newton &2001-01-05 &0111190701 & 50.7 & 608/257/200 & 1.99 & $2.65_{-0.20}^{+0.21}$ & $10.02_{-0.52}^{+1.18}$  \\
135241.94+334138.5 & 1.514 & Chandra & 2004-06-05&4867  & 31.1 & 242 & 1.60 & $2.04_{-0.20}^{+0.20}$ & $7.19_{-0.64}^{+0.86}$  \\
141325.92+435146.5 & 1.526 & XMM-Newton &2002-07-10& 0103660101 & 21.4 & 217/130/106 & 0.99 & $2.19_{-0.25}^{+0.30}$ & $3.46_{-0.43}^{+0.61}$  \\
145457.11+585804.0 & 2.180   & XMM-Newton &2016-06-29& 0783881301 & 16.7 & 354/131/128 & 1.10 & $2.15_{-0.14}^{+0.15}$ & $12.39_{-0.62}^{+1.39}$  \\
162710.37+350117.9 & 2.290   & XMM-Newton &2012-01-17& 0674811001 & 8.1 & 128/71/81 & 1.42 & $2.21_{-0.28}^{+0.31}$ & $16.89_{-2.31}^{+3.33}$ \\
\enddata
\tablecomments{Column (1): Name of the object. Column (2): Redshift derived from the SDSS DR16 quasar catalog. Columns (3), (4) and (5): X-ray observatory, observation date, and observation ID. Column (6): Cleaned exposure time. Column (7): Net source counts in certain bands like Table~\ref{tbl:obs}. Column (8): Galactic neutral hydrogen column density. Column (9): Hard X-ray photon index. Column (10): Luminosity in the \hbox{rest-frame} \hbox{2--10~keV} band, corrected for the Galactic absorption.}
\label{tab:x_ray_12}
\end{deluxetable*}

% F7 1643 spectral

\end{document}